\documentclass[12pt]{article}
\usepackage[a4paper , hmargin=2cm, vmargin=2.5 cm]{geometry}
\usepackage[utf8]{inputenc}
\usepackage{graphicx}
\usepackage{hyperref}
\usepackage{latexsym, verbatim,amsmath,bm}
\usepackage{amsfonts}
\usepackage{xcolor,soul}
\sethlcolor{yellow}

\usepackage{caption} 

\title{On the competition for ultimately stiff and strong architected materials}%
\author{Morten N. Andersen, Fengwen Wang and Ole Sigmund \\
    Department of Mechanical Engineering, Solid Mechanics, \\
    Building 404, Technical University of Denmark,\\ 					DK-2800, Lyngby, Denmark\\
    Email: morand@mek.dtu.dk}%

\date{\today}

\newcommand{\be}{\begin{equation}}
\newcommand{\ee}{\end{equation}}
\newcommand{\bea}{\begin{eqnarray}}
\newcommand{\eea}{\end{eqnarray}}
\newcommand{\beano}{\begin{eqnarray*}}
\newcommand{\eeano}{\end{eqnarray*}}

\begin{document}

\maketitle

\begin{abstract}
Advances in manufacturing techniques may now realize virtually any imaginable microstructures, paving the way for architected materials with properties beyond those found in nature. This has lead to a quest for closing gaps in property-space by carefully designed metamaterials. Development of mechanical metamaterials has gone from open truss lattice structures to closed plate lattice structures with stiffness close to theoretical bounds. However, the quest for optimally stiff and strong materials is complex. Plate lattice structures have higher stiffness and (yield) strength but are prone to buckling at low volume fractions. Hence here, truss lattice structures may still be optimal. To make things more complicated, hollow trusses or structural hierarchy bring closed-walled microstructures back in the competition. Based on analytical and numerical studies of common microstructures from the literature, we provide higher order interpolation schemes for their effective stiffness and (buckling) strength. Furthermore, we provide a case study based on multi-property Ashby charts for weight-optimal porous beams under bending, that demonstrates the intricate interplay between structure and microarchitecture that plays the key role in the design of ultimate load carrying structures. The provided interpolation schemes may also be used to account for microstructural yield and buckling in multiscale design optimization schemes.
\end{abstract}


\newpage

\section{Introduction} \label{sec:intro}

Stiffness measures a structure's ability to resist deformation when subjected to external load. Strength measures the ultimate load carrying capability of a structure. A structure can be stiff but have low strength, such as a longitudinally compressed slender steel rod that is initially stiff but looses stability even for small loads. Oppositely, a structure can have high strength but low stiffness, such as a grass straw swaying in the wind. Engineering structures must be both stiff and strong. Bridge decks or airplane wings are only allowed to deflect a certain amount and at the same time they must be able to withstand substantial forces. Another engineering goal is to minimize structural mass and material consumption, partly to save weight and thereby fuel consumption in moving structures, and partly to save money and natural resources in the manufacturing process. Structural optimization can be performed on the macro-scale based on available materials or it can be performed on nano- or micro-scale by looking for improved material alloys or by taking existing base materials and tailoring their microstructures to obtain certain functionalities. The latter constitutes a very hot research topic and goes under many names, such as architected materials, metamaterials, tailored materials, microstructured materials, etc. and is the subject of the present work.

A "stiff competition" \cite{Mil18} for architected materials has been going on for decades. Already in the 1980's, applied mathematicians found that microstructures meeting the upper Hashin-Shtrikman bounds \cite{HasSht63} can be realized by so-called rank-$n$ laminates \cite{FraMur86,Mil86,Ave87-01}. These, however, are physically unrealistic since they require laminations at up to $n=6$ widely differing length-scales, although they in the low volume fraction case can be simplified to one length scale \cite{Chr86}. Importantly, these optimally stiff microstructures are closed-walled, from now on denoted plate lattice structures (PLS). For intermediate volume fractions, PLS that are optimal in the low volume fraction limit may be thickness-scaled to yield practically realizable microstructures with Young's moduli within 10\% of the theoretically achievable values \cite{SigAagAnd16,BerWadMcm17,TraSigGro18}. If for reasons like manufacturability or permeability, one is restricted to open truss lattice structures (TLS), this comes at the cost of an up to three-fold decrease in attainable stiffnesses \cite{Chr86,SigAagAnd16,BerWadMcm17,WanGroSig18}.

The advent of advanced manufacturing techniques at micro- and nano-scale has also resulted in a "strength competition". Partly due to manufacturing challenges, contestants have mainly been open-walled TLS or hollow truss lattice structures (hTLS) \cite{MezDasGre14,BauSchKra16,SchJacCar11}. Here, the saying ``smaller is stronger'' becomes relevant as base material yield strength grows with decreasing scale due to less chances of defects at the nano-scale \cite{ZhaVyaLi19,PorWidKoc20}. Hence, microstructures with remarkable strength and resilience have been realized, culminating with recent PLS nano-structures making use of compressive ultimate strength of up to 7 GPa \cite{CroBauVal20} attainable by sub-micro scale beams made from Pyrolyzed Carbon. Very few research groups, however, have considered strength optimization in terms of first onset of microstructural stability or buckling. Two recent exceptions report high buckling strength for PLS at higher volume fractions \cite{CroBauVal20,TanDiaNoh18}.

Considering the recent reports on high strength and stiffness \cite{CroBauVal20,TanDiaNoh18} of plate lattice structures (PLS), it may be tempting to conclude that the combined competition for stiffness and strength is over and the winner is PLS. However, this is not the case. PLS are only optimal under certain conditions as we will demonstrate. For lower volume fractions, the walls in PLS become thin and unstable and thicker and more stable truss lattice structures (TLS or hTLS) take over. This conclusion, however, cannot be drawn from simple studies of (specific) stiffness-strength diagrams but requires case studies. Here we will draw such conclusions based on a simple square cross-sectioned beam in bending. Other case studies may lead to other conclusions. For example, we show that if the beam has variable width and fixed height, no microstructure is ever optimal. Here, the solid beam always provides the stiffest and strongest solution.

Previously, macroscopic and microscopic instabilities have been investigated in 2D for random and periodic porous elastomers under large deformation \cite{TriNesSch06,MicLopCas07}. Macroscopic instability of 3D random porous elastomers has been studied using second order homogenization assuming linear comparison composites, where macroscopic instability is identified by the loss of strong ellipticity of the homogenized constitutive  model. \cite{LopCas07a,LopCas07b}. A systematic study of microscopic buckling for 3D architected materials, has to our knowledge, not been performed before.

In this study, we focus on elastic microstructures and do not account for material non-linearities but identify microstructural strength by first onset of local yield or elastic instability - whichever happens first. We only study stretch-dominated microstructures, which are known to provide optimal or near-optimal stiffness. However, discussions; developed interpolation schemes; as well as methods for determining optimality for certain applications; are general and apply to any other microstructures, albeit with lower obtainable stiffnesses. Also, we limit ourselves to cubic symmetric or isotropic microstructures due to their general applicability and stability to varying load situations, although we know that anisotropic microstructures like transverse honeycombs may perform much better for specific and well-defined load scenarios \cite{WanGroSig18,Bit97}. Again, however, methods and conclusions developed will also apply to any anisotropic materials.

Apart from providing new insights in stiffness and strength of extremal microstructures, the results of our study has a number of other potential applications and implications. First, the computed effective stiffness and strength properties may directly be used in the modelling and evaluation of lattice and infill structures realized by additive manufacturing techniques. Describing the implicit CAD geometry of periodic lattice structures is a tedious task and subsequent meshing quickly results in huge and unmanageable finite element models. Therefore, simple material interpolation laws that provide stiffness as well as strength estimates for specific microstructures as function of filling fraction are in high demand and have yet to be performed. Second, the same interpolation schemes may directly be used as material interpolation functions in multiscale structural topology optimization problems \cite{GroStuSig19}. Hitherto, such multiscale topology optimization approaches have almost entirely focussed on pure linear stiffness optimization ignoring possible microstructural failure mechanisms. Our results pave the way for including both yield and local stability constraints in such schemes with manageable computational overhead.

The paper is composed as follows. First we list and discuss existing theoretical bounds on microstructural stiffness and yield strength. Next, we propose to use two-term interpolation schemes for material stiffness, buckling and yield strength to improve on existing one-term schemes for up to moderate volume fractions and discuss their extensions to hierarchical microstructures. For a number of commonly used isotropic and cubic symmetric microstructures from the literature we perform analytical and extensive numerical evaluations to provide coefficients for their associated two-term stiffness and strength interpolation schemes. Finally, we provide a beam example that demonstrates the use of our material interpolation laws and highlights how truss and plate lattice structures take turns in being optimal, depending on beam span and base material properties, even for this simple case study.

\section{Theoretical bounds} \label{sec:bounds}

\begin{figure}[!t]
	\begin{center}
		\includegraphics[width=0.6\textwidth]{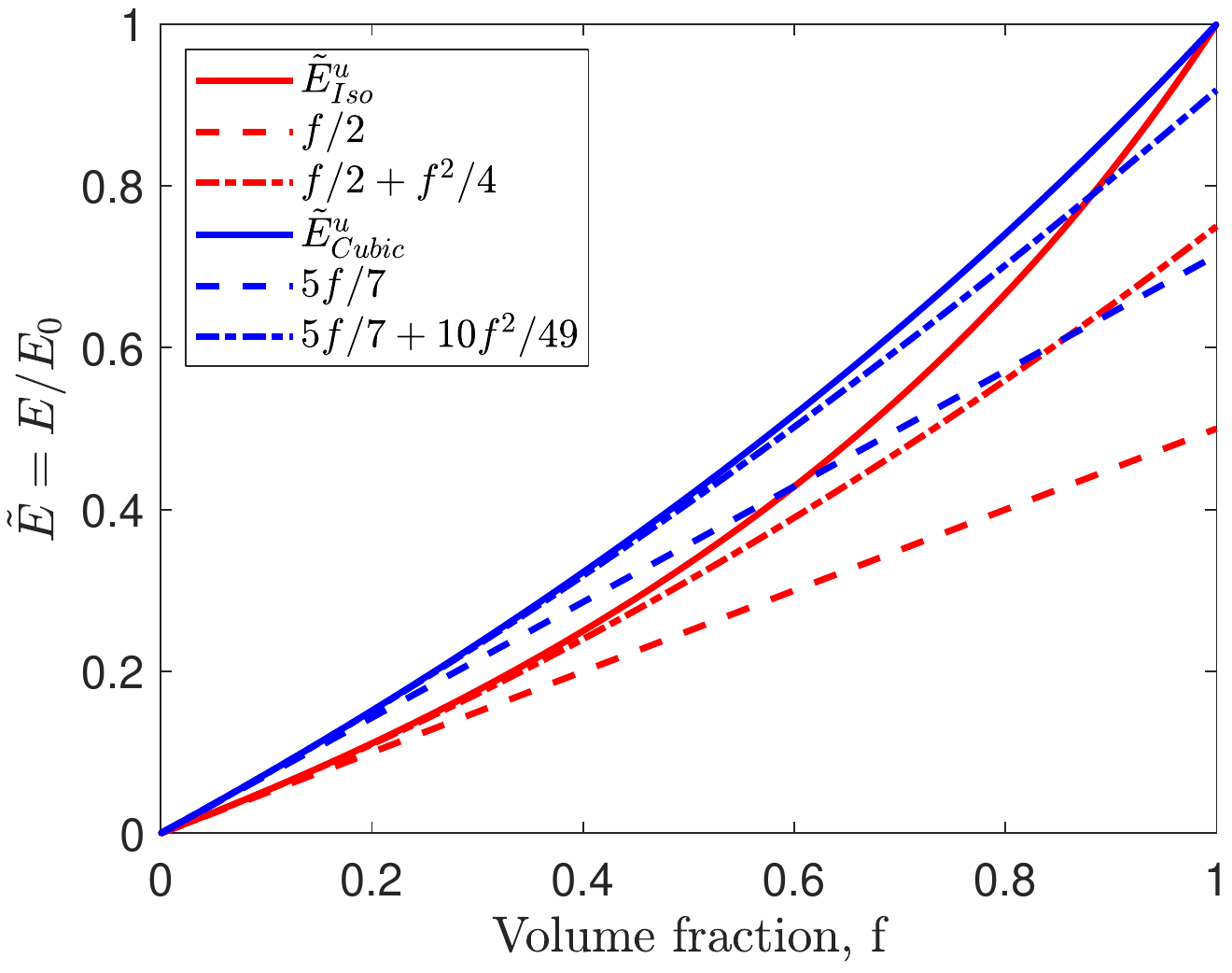}
	\end{center}
	\caption{Plot of Hashin-Shtrikman bounds given by (\ref{eq:HS}) for isotropic and cubic symmetric microstructures and their first and second order approximations.}  \label{Fig:Bounds}
\end{figure}

Hashin-Shtrikman (HS) bounds provide upper limits on attainable Young's moduli for porous microstructures \cite{HasSht63}. These are rather complex expressions given in terms of base material properties: Poisson's ratio $\nu_0$ and Young's modulus $E_0$, as well as volume fraction $f$ (see Appendix \ref{App:bounds}). However, variability in terms of Poisson's ratio is small in the range of usual (compressible) base material values $\nu_0 \in [0, 1/2[$, hence selecting a value of $\nu_0=1/3$ at most gives an error of 1\% in aforementioned interval. With this assumption, HS bounds for isotropic and cubic symmetric materials simply become

\be
   E^u_{Iso} = \frac{f}{2-f} E_0, \ {\rm and} \ E^u_{Cubic} = \frac{5f}{7-2f} E_0. \label{eq:HS}
\ee
From these two bounds and their graphs in Figure \ref{Fig:Bounds}, it is clear that requiring isotropy over simpler cubic symmetry deteriorates attainable stiffness with up to a factor of $10/7$ (43\%) for low volume fractions.

An expression for a yield strength bound for uni-axial loading of isotropic microstructures was derived by Casta\~{n}eda \cite{Cas91} (however, often attributed to Suquet \cite{Suq93}) and only depends on the volume fraction and yield (or ultimate) stress limit of the base material $\sigma_0$, i.e.
\be
   \sigma^u_{y} = \frac{2 f}{\sqrt{4+11/3 (1-f)}} \sigma_0 = \frac{6 f}{\sqrt{69 - 33f}} \sigma_0. \label{eq:Suquet}
\ee

An assumption behind the Casta\~{n}eda bound is that it ignores stress concentrations and hence approaches the solid material yield strength $\sigma_0$ as volume fraction approaches one. At first thought, this may seem logical but this is actually not physically possible. As volume fraction approaches one, voids approach zero size. Any small void will cause stress concentrations and hence yield strength of the porous material will not approach that of the solid for vanishing hole size but actually be lower by some stress concentration factor. For a small spherical void, this stress concentration factor is 2 for uni-axial loading. Hence, the yield bound is not optimal in the sense that there exist no physical microstructures that can achieve it. This is especially pronounced for higher volume fractions. Hence, a simplified bound that takes some of this stress concentration at higher volume fractions into account could be the linear function $\sigma_y^u = \frac{\partial \sigma_y^u}{\partial f}\Bigr\rvert_{f = 0} f \sigma_0 = \frac{2\sqrt{69}}{23}f\sigma_0 \approx \frac{13}{18} f\sigma_0$.

\section{Material interpolation schemes} \label{sec:interpolations}

Literature often reports derived, computed or measured porous material properties in terms of simple, single order polynomials \cite{GibAsh97}. These may be sufficiently accurate in the very low volume fraction limit or for smaller subintervals. However, first order polynomials are insufficient for larger volume fractions and more general studies. As will be shown later, errors can be huge even for quite low volume fractions. Inferiority of one-order v.s. two-order polynomial expansions of the HS bounds are for now already visible in Figure \ref{Fig:Bounds}. Hence, we here compute and list all material properties as two-term polynomials that make them valid up to volume fractions of 0.5.

Assuming stretching-dominated microstructures, two-term polynomials approximations for effective Young's modulus, buckling and yield strength are
\bea
     E_f&=&E(f)=\tilde{E}(f) \ E_0 = (a_0 f + a_1 f^2) E_0, \label{eq:stiff} \\
    \sigma_{c,f}&=&\sigma_c(f)=
            \tilde{\sigma}_c(f) E_0 = (b_0 f^{n_0} + b_1 f^{n_0+1}) E_0, \label{eq:buck} \\
    \sigma_{y,f}&=&\sigma_y(f)=
     \tilde{\sigma}_y(f) \sigma_{0} = (c_0 f + c_1 f^2) \sigma_0  \label{eq:yield}
\eea
where all coefficients ($0 <(a_i, b_i, c_i)<1$) and exponent $n_0$ are estimated from analytical and/or numerical studies (see Table \ref{tab:uniaxial} for numbers for specific microstructures and Section \ref{sec:numerics} for numerical details of their derivation). As it will turn out, $n_0=2$ for TLS and $n_0=3$ for PLS, which means that the buckling strengths of the two kinds of microstructures are notably different in terms of volume fraction dependence\footnote{For a fixed beam length or plate dimensions, beam buckling depends on cross-sectional area (and hence volume fraction) squared and plate buckling depends on thickness (and hence volume fraction) cubed.}. These exponents will later be shown to be decisive factors when looking for the optimal microstructural material morphology for a given application.

We define the effective strength of a porous material in compression as the minimum of buckling strength $\sigma_{c,f}$ (\ref{eq:buck}) and yield strength $\sigma_{y,f}$ (\ref{eq:yield}). Since polynomial order of the former always is higher than for the latter, buckling strength will always be the decisive one for lower volume fractions as also intuitively expected. The transition to yield controlled failure depends on considered micro-architecture.

\subsection{Interpolation schemes for hierarchical microstructures}

Lakes (1993) derived expressions for Young's modulus and buckling strength of $n$'th order architected microstructures based on the commonly used one-term polynomial material interpolation functions. Assuming self-similar hierarchy, i.e. each level has the same microstructure and volume fraction, the interpolation functions for Young's modulus, buckling and yield strengths of an $n$'th order stretch-dominated hierarchical microstructure are
\bea
     E_{f,n} &=& E_n(f) = a_0^n f E_0, \label{eq:Lakes-stiff} \\
    \sigma_{c,f,n}&=&\sigma_{c,n}(f) = b_0 a_0^{n-1} f^{1+\frac{n_0-1}{n}} E_0, \label{eq:Lakes-buck} \\     \sigma_{y,f}&=&\sigma_y(f)= c_0^n f \sigma_0, \quad n \in \mathbb{Z}^+. \label{eq:Lakes-yield}
\eea
Note here that Lakes' paper used $n$ instead of $n-1$ for the exponent on $a_0$ in  (\ref{eq:Lakes-buck}), which is a typo. (Re)derivations for all three expressions as well extension to the more practical two-term scheme (\ref{eq:stiff})-(\ref{eq:yield}) as well as fully general interpolation schemes can be found in Appendix \ref{App:hierarchical}.

There are two important remarks to these expressions. First, we note that $a_0$, $b_0$ and $c_0$ in (\ref{eq:stiff})-(\ref{eq:yield}) always are (sometimes significantly) smaller than one and hence hierarchy ($n>1$) will inevitably decrease performance of all material properties for a given volume fraction $f$. For the buckling strength case, however, hierarchical order higher than one may still be an advantage if the volume fraction is low enough. In particular, if the volume fraction is lower than
\be
    f_{lim}= a_0^\frac{n}{n_0-1} \ {\rm for} \ n \geq 2. \label{eq:volfrac-lim}
\ee
For TLS ($n_0=2$), this means that second order hierarchical microstructure is advantageous with respect to buckling strength for volume fractions smaller than $a_0^2$. Similarly, for PLS ($n_0=3$), second order hierarchy is advantageous for volume fractions smaller than $a_0$, i.e. for higher volume fraction than for TLS.

Second, we note that for TLS ($n_0=2$), the volume fraction exponent in (\ref{eq:Lakes-buck}) is $1+\frac{n_0-1}{n}=1+\frac{1}{n}$. This means that strength depends on the volume fraction to the $3/2$ power for a second order hierarchical structure. Hence, the dependence is not first order as sometimes claimed in the literature. Only in the limit of infinite order does the dependence converge to first order. However, at the same time the factor $a_0^{n-1}$ in (\ref{eq:Lakes-buck}) would go to zero and thus nothing would be gained from this linear dependence! For PLS, ($n_0=3$), the volume fraction exponent in (\ref{eq:Lakes-buck}) is $1+\frac{n_0-1}{n}=1+\frac{2}{n}$. Hence here, strength dependence on volume fraction is raised to power $2$ for a second order hierarchical structure, which makes it depend on $f$ the same way as the first order hierarchical TLS, at the cost of a buckling strength reduction factor of $a_0$. When, as it turns out, $a_0$ is much bigger for PLS than TLS, this suddenly makes the second order hierarchical PLS very attractive compared to the first order TLS (see actual numbers later).

\section{Microstructures} \label{sec:numerics}

\begin{figure}[!htb]
	\begin{center}
		\includegraphics[width=1\textwidth]{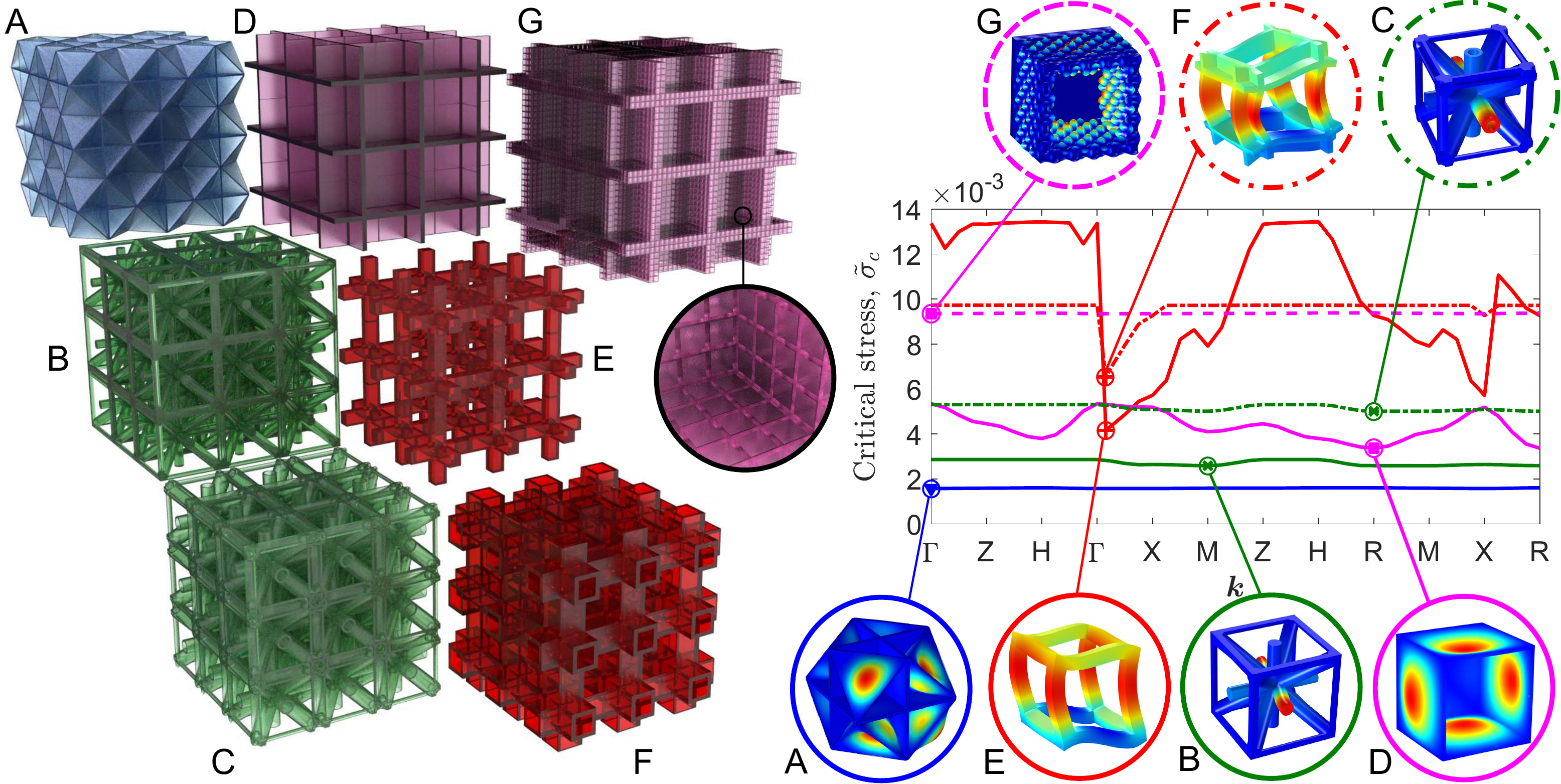}
	\end{center}
	\caption{Buckling strength of microstructures. Illustrations of considered geometries for volume fraction $f\approx 0.2$. A: Iso-PLS, B: Iso-TLS, C: Iso-hTLS, D: SC-PLS, E: SC-TLS, F: SC-hTLS and G: a second order hierarchical version of D. Right half shows lowest buckling value band over the edges of the irreducible Brillouin zone and associated worst case buckling modes for the considered microstructures. Each band is based on 33 unique evaluation points, except for the second order hierarchical SC-PLS, G, which is based on 6 unique evaluation points. Points marked by circles indicate worst case critical buckling stress for each microstructure.}  \label{Fig:Geom}
\end{figure}

As representatives of near-optimal, isotropic and cubic symmetric truss (TLS) and plate (PLS) elastic microstructures, we choose the six illustrated in Figure \ref{Fig:Geom}A-F. We consider three simple cubic (SC) microstructures composed of flat plates (D: SC-PLS), bars with square cross sections (E: SC-TLS) and its hollow bar counterpart (F: SC-hTLS), respectively. Similarly, we consider three isotropic microstructures synthesized by combination of SC and body-centered cubic lattice with thickness ratio between SC and BCC plates fixed to $t_{SC}/t_{BCC}=8 \sqrt{3}/9$ \footnote{The thickness changes to $t_{SC}/t_{BCC}=\sqrt{3}$ for higher volume fractions to maintain isotropy \cite{BerWadMcm17}.} and the area ratio between the two circular bar groups of the corresponding TLS fixed to $A_{SC}/A_{BCC}=4/3\sqrt{3}$. These three isotropic lattice structures, are hereafter referred to as (A: Iso-PLS), (B: Iso-TLS) and the hollow version (C: Iso-hTLS), respectively. The Iso-PLS has near optimal Young's modulus (\cite{GurDur14,BerWadMcm17,TanDiaNoh18}), only a few percent inferior to the bound (\ref{eq:HS}) for moderate volume fractions. The six micro architectures are color-coded as SC-TLS (red), SC-hTLS (dash-dotted red), SC-PLS (magenta), Iso-TLS (green), Iso-hTLS (dash-dotted green) and Iso-PLS (blue). This color scheme will be used also for coloring of graphs throughout this work with hierarchical versions using same colors but dashed curves instead of solid.

The hollow versions of the two TLS, are inspired by \cite{ZheSmiSpa16}, however with the material inside bar crossings maintained for simplicity, stiffness and stability. Considering microstructures with hollow crossings will significantly deteriorate stiffness and are hence left out of this study. The thicknesses of the hollow bars are tailored to avoid wall-buckling \cite{WeaAsh97} within the volume fraction range of interest, $f \in [10^{-4}, 0.5]$ and hence to maintain the similar critical buckling modes as their solid counterparts. Detailed expressions for the hollow cross-section dimensions are listed in Appendix \ref{App:hollow}.

Figure \ref{Fig:Geom}G also shows a hierarchical ($n=2$) version of the SC-PLS structure to be discussed later.

\section{Modelling} \label{sec:numerics}

Effective properties of the considered microstructures are computed using analytical studies as well as numerical homogenization and finite element analyses. PLS are discretized by shell elements in the low volume fraction range ($f < 10^{-2}$), while TLS are modelled analytically. Continuum elements are used for both cases at higher volume fractions up to 0.5. In this work, microstructural failure is defined as the onset of yielding or buckling and hence we do not investigate large deformation, crushing or other non-linear behaviour. Furthermore, the microstructures are modelled as ideal, i.e. with no imperfections and the base material is linear elastic, homogenous and isotropic. Yield strength is computed as the product of the applied macroscopic stress and the ratio between  the yield stress of the base material and the maximum von Mises stress for the cell. The maximum von Mises stress is computed at the center of plates or struts making up the microstructures and thus does not take local stress concentrations into account.

The numerical computation of buckling strength is quite elaborate and has to our knowledge not been performed before for 3D microstructures. The same macroscopic stress state as used for calculating yield strength forms the basis for a linear buckling analysis based on Floquet-Bloch wave theory (\cite{TriSch93,NevSigBen01,ThoWanSig17}). By searching over the wave-vector space spanned by the edges of the irreducible Brillouin zone, we identify the most critical load value over all possible wavelengths and mode directions. Herein, a small or large wavelength, compared to the unit cell size, corresponds to microscopic or macroscopic instability, respectively. Hence, we identify the most critical mode amongst all modes ranging from local to global.

A resulting band diagram for the considered microstructures for $f\approx 0.2$ is shown in Figure \ref{Fig:Geom}, where smallest value over all wave-vectors for each microstructure represents its critical buckling stress. Here it is clearly seen how the isotropic microstructures (blue PLS and green TLS and hTLS) have almost direction and wavelength independent critical buckling spectra but, at least for the PLS and TLS, inferior buckling strengths. On the other hand, the cubic symmetric microstructures (magenta PLS and red TLS and hTLS) have more anisotropic but nevertheless superior buckling responses. Inserts in circles show the most critical buckling modes  for each microstructure. Critical modes for the SC-TLS and hTLS microstructures are global shear failure modes (just right of the $\Gamma$ point), whereas critical modes for all other microstructures are local, either cell-periodic or cell anti-periodic. Buckling instability for the second order hierarchical SC-PLS (magenta dashed curve) is independent on wave number and is associated with cell wall buckling at the lower hierarchical level.

\begin{table}[!b]
\begin{center}
\caption*{Polynomial material interpolation coefficients}
\begin{tabular}{|l|c|c|c|c|c|c|}
\hline
     & $\tilde{E}$  & $n_0$ & $\tilde{\sigma}_c$  & $\tilde{\sigma}_y$  \\ \hline
Iso-PLS            & \begin{tabular}[c]{@{}c@{}}$a_0=1/2=0.5$\\ $a_1=0.228$\end{tabular}           & 3     & \begin{tabular}[c]{@{}c@{}}$b_0=0.200$\\ $b_1=0.184$\end{tabular} & \begin{tabular}[c]{@{}c@{}}$c_0=\frac{16\sqrt{111}}{333}\approx 0.506$\\ $c_1=0.252$\end{tabular}   \\ \hline
Iso-TLS            & \begin{tabular}[c]{@{}c@{}}$a_0=1/6\approx 0.167$\\ $a_1=0.464$\end{tabular}             & 2     & \begin{tabular}[c]{@{}c@{}}$b_0= \frac{\pi}{90}\approx 0.035$\\ $b_1=0.143$\end{tabular} & \begin{tabular}[c]{@{}c@{}}$c_0=1/6\approx 0.167$\\ $c_1=0.284$\end{tabular}   \\ \hline
Iso-hTLS           & \begin{tabular}[c]{@{}c@{}}$a_0=1/6\approx 0.167$\\ $a_1=0.589$\end{tabular}           & 3/2     & \begin{tabular}[c]{@{}c@{}}$b_0=\frac{\sqrt{0.45\pi}}{30\sqrt[4]{0.96}}\approx 0.040$\\ $b_1=0.089$\end{tabular} & \begin{tabular}[c]{@{}c@{}}$c_0=1/6\approx 0.167$\\ $c_1=0.345$\end{tabular}   \\ \hline
SC-PLS             & \begin{tabular}[c]{@{}c@{}}$a_0=5/7\approx 0.714$\\ $a_1=0.147$\end{tabular}             & 3     & \begin{tabular}[c]{@{}c@{}}$b_0=0.350$\\ $b_1=0.229$\end{tabular} & \begin{tabular}[c]{@{}c@{}}$c_0=\frac{10\sqrt{21}}{63}\approx 0.727$\\ $c_1=0.117$\end{tabular}   \\ \hline
SC-TLS             & \begin{tabular}[c]{@{}c@{}}$a_0=1/3\approx 0.333$\\ $a_1=0.517$\end{tabular}             & 2     & \begin{tabular}[c]{@{}c@{}}$b_0=\frac{6}{108}\approx 0.056$\\ $b_1=0.196$\end{tabular} & \begin{tabular}[c]{@{}c@{}}$c_0=1/3\approx 0.333$\\ $c_1=0.400$\end{tabular}   \\ \hline
SC-hTLS            & \begin{tabular}[c]{@{}c@{}}$a_0=1/3\approx 0.333$\\ $a_1=0.663$\end{tabular}           & 5/3     & \begin{tabular}[c]{@{}c@{}}$b_0=\frac{1}{6\sqrt[3]{5}} \approx  0.098$\\ $b_1=0.043$\end{tabular} & \begin{tabular}[c]{@{}c@{}}$c_0=1/3\approx 0.333$\\ $c_1=0.520$\end{tabular}   \\ \hline
Bounds (isotropic) & \begin{tabular}[c]{@{}c@{}}$a_0=1/2=0.5$\\ $a_1=1/4=0.25$\end{tabular}   & -     & - & \begin{tabular}[c]{@{}c@{}}$c_0=\frac{2\sqrt{69}}{23}\approx 0.722$\\ $c_1=\frac{11\sqrt{69}}{529}\approx 0.173$\end{tabular} \\ \hline
Bounds (cubic)     & \begin{tabular}[c]{@{}c@{}}$a_0=5/7\approx 0.714$\\ $a_1=10/49\approx 0.204$\end{tabular} & -     & - & \begin{tabular}[c]{@{}c@{}} - \end{tabular} \\ \hline
\end{tabular}%
\caption{Analytically and numerically derived coefficients for two-term polynomial interpolation schemes proposed in Section 3. Coefficients given as fractions are based on analytical studies in the low volume fraction limit. Other coefficients are based on numerical studies.}
\label{tab:uniaxial}
\end{center}
\end{table}

The best performing solid microstructure with respect to buckling strength for $f\approx 0.2$ is the red SC-TLS with its worst case global shear mode, right next to the $\Gamma$-point exhibiting the highest critical stress value over the solid microstructures (red SC-TLS and magenta -PLS, green Iso-TLS and blue -PLS). This latter case corresponds well to analytical studies from the literature, c.f. \cite{HagPapVaz14,BluSigPou20}. The plot also shows that the hollow microstructures red dash-dotted SC-hTLS and green dash-dotted Iso-hTLS, as well as the magenta dashed hierarchical microstructure, perform better than their solid counterparts. More discussions follow later and numerical details are given in Appendix \ref{App:numerics}.

Following these extensive analytical and numerical analyses, material interpolation coefficients $a_0$, $b_0$ and $c_0$ for the interpolations schemes proposed in Section 3 are determined from analytical add-up models (checked with truss FE model) for TLS and numerically using shell finite elements for PLS in the low volume fraction range (see Appendix \ref{App:hierarchical} for details). The second order terms $a_1$, $b_1$ and $c_1$ are determined from curve fits of remaining data points (a total of 16 volume fractions for each microstructure provide the basis for the interpolations) based on a continuum FE model. The resulting coefficients are listed in Table \ref{tab:uniaxial}. All numbers given as fractions are analytically obtained values. The justification for a two-term interpolation function can be trivially verified by inserting data from Table \ref{tab:uniaxial} into (\ref{eq:stiff})-(\ref{eq:yield}). The difference between one-term and two-term scales linearly with volume fraction. For example, using data for the Iso-TLS structure and volume fraction $f=0.2$, the usual one-term interpolation scheme underestimates stiffness with $e=\big(\tilde{E}_f\left(a_0,a_1\right)-\tilde{E}_f\left(a_0\right)\big) / \tilde{E}_f\left(a_0\right) = 56\%$ and analogously for buckling stress, an underestimation of 82\%.

A lot can be learned from studying Table \ref{tab:uniaxial} in detail. First we, as expected, observe that PLS reach the upper bounds on Young's modulus for isotropic and simple cubic microstructures for low volume fractions. The same two microstructures have stresses very close to the yield bound (\ref{eq:Suquet}). On the other hand, the PLS are, at least for lower volume fractions, suboptimal with respect to buckling stability, since their power $n_0=3$ is higher than the TLS ($n_0=2$). Interestingly, the hTLS also beat the TLS with isotropic and cubic symmetric stability exponents of $n_0=3/2$ and $5/3$, respectively. These observations will be discussed deeper later.

\begin{figure}[!t]
	\begin{center}
		\includegraphics[width=17cm]{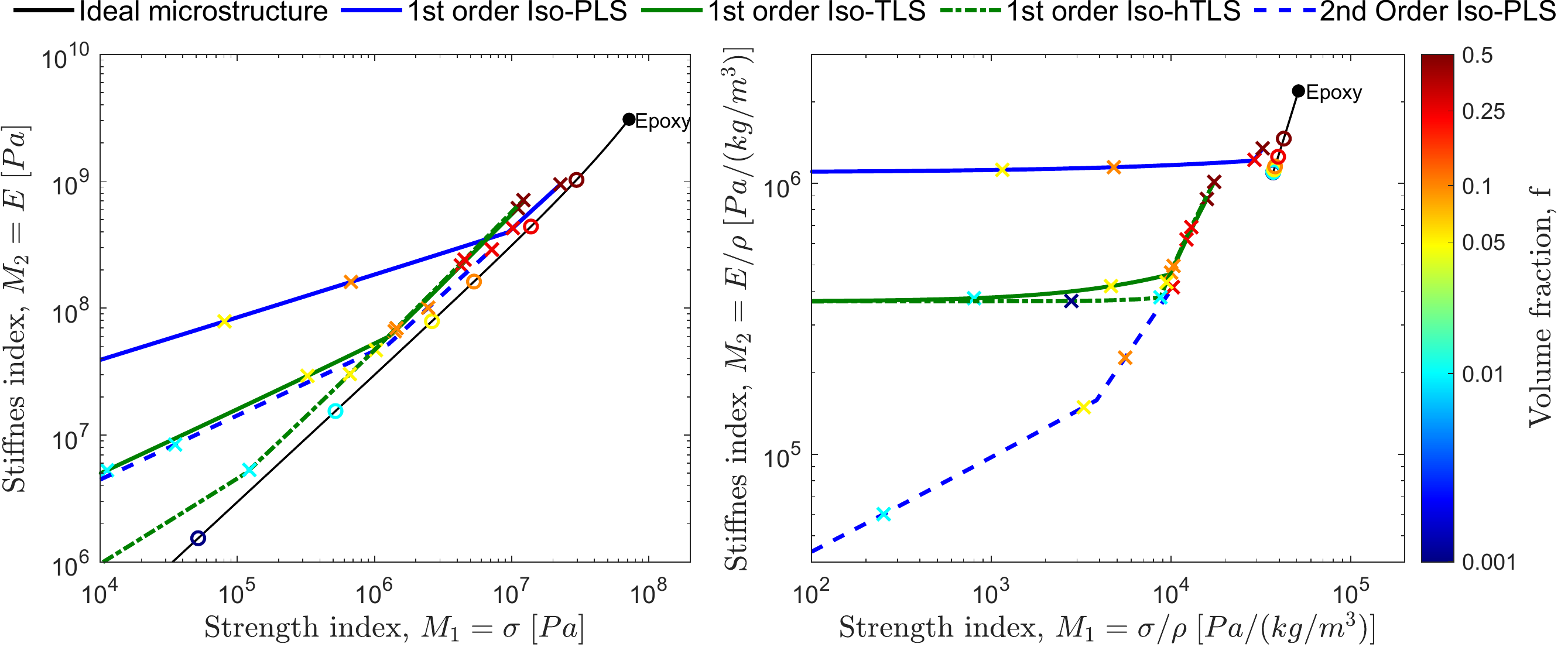}
	\end{center}
	\caption{Strength versus stiffness for selected microstructures. Strength-Stiffness (left) and Specific Strength-Stiffness (right) plots for isotropic first and second order hierarchical as well as the isotropic hollow hTLS. Blue lines indicate PLS and green TLS. Solid lines indicate simple and dashed second order hierarchical microstructures. The dash-dotted green lines indicate the hTLS microstructure and the ideal material performance (reaching both stiffness and yield bounds) is indicated by the black curves. The colored markers indicate volume fractions according to the color bar.}  \label{Fig:StiffnessStrength}
\end{figure}

Using Epoxy as base material (see Table \ref{tab:materialprops} for material properties) for isotropic first and second order hierarchical  microstructures as well as hTLS, obtainable properties are plotted in strength-stiffness and specific strength-stiffness Ashby plots in Figure \ref{Fig:StiffnessStrength}. Full and dashed blue lines indicate PLS properties for varying volume fractions for first and second order hierarchy, respectively. Green lines indicate the same for TLS properties. Finally, the dash-dotted green line indicates properties of the hTLS. Colored crosses indicate points for specific volume fractions, starting with red at $f=0.5$ and descending to blue at $f=0.001$. Kinks in property lines indicate transition points from yield controlled (higher volume fractions) to buckling controlled (lower volume fractions). Black curves indicate the properties of the ``ideal microstructure'' that simultaneously attains the stiffness and yield bounds. Again, colored markers indicate volume fractions. Without the availability of cross-property bounds that relate strength and stiffness, the tightest property bounds for a given volume fraction are given by horizontal or vertical lines, extending from the colored volume fraction markers. From both strength-stiffness as well as specific strength-stiffness plots, it is clear that one cannot beat the material properties of the base material (here Epoxy), no matter what microstructure or hierarchical level is used. Hence, from these graphs it is not obvious why one would consider using porous microstructure at all. One needs to study specific applications to come up with an answer to this question.

\section{Beam model}

We seek a simple engineering design problem that may benefit from high stiffness, high strength microstructures and illustrates the role of different microstructural effects and properties. The simplest imaginable structure for this purpose is the mass minimization of a beam in bending.

A simply supported Bernoulli-Euler beam with rectangular cross-section and width $w$, height $h$ and length $L$ is subject to equal but oppositely oriented bending moments $V$ at both ends, hence its moment distribution is constant and shear stresses are zero. Mass, mid-span displacement and maximum stress of the beam are
\be
   m = f \rho_0 L w h, \quad
   \delta = \frac{3}{2} \frac{V L^2}{E_f w h^3} \quad {\rm and} \quad
   \sigma_{max} = \frac{6 V}{w h^2}. \label{eq:beamprops}
\ee
Now we want to minimize the mass of this beam subject to a displacement constraint $\delta^*$ and avoidance of yield and microstructural buckling. From now on, we assume a variable square cross-section ($w=h$) but variable width or height cases as well as details on their derivations are included in Appendix \ref{App:beam} and discussed later.

The minimal mass of the beam subject to a displacement constraint $\delta^*$ is
\be
   m_\delta = \sqrt{\frac{3}{2}} \sqrt{\frac{V}{\delta^*}} L^2 \frac{1}{M_2}, \quad
   M_2 = \frac{\sqrt{E_f}}{f \rho_0} = \psi_B^\delta \frac{\sqrt{E_0}}{\rho_0}, \quad
   \psi_B^\delta = \frac{\sqrt{\tilde{E}_f}}{f}, \label{eq:m_delta_square}
\ee
where $M_2$ is the \emph{material stiffness index} and $\psi_B^\delta$ is the \emph{microscopic shape factor for elastic bending} (\cite{Ash99}) to be maximized in order to decrease beam mass (see further details in Appendix \ref{App:beam}).

The mass subject to yield or buckling failure constraints is given by the maximum of the masses subject to either local microscale buckling constraint
\be
   m_c = (6 V)^{\frac{2}{3}} L \frac{1}{M_1}, \quad
   M_1 = \frac{\sigma_{c,f}^{\frac{2}{3}}}{f \rho_0} = \psi_B^c \frac{E_0^{\frac{2}{3}}}{\rho_0}, \quad
   \psi_B^c = \frac{{\tilde{\sigma}_{c,f}}^{\frac{2}{3}}}{f}
   \label{eq:m_sigma_square}
\ee
or yield constraint
\be
   m_y = (6 V)^{\frac{2}{3}} L \frac{1}{M_1}, \quad
   M_1 = \frac{\sigma_{y,f}^{\frac{2}{3}}}{f \rho_0} = \psi_B^y \frac{\sigma_0^{\frac{2}{3}}}{\rho_0}, \quad
   \psi_B^y = \frac{\tilde{\sigma}_{y,f}^{\frac{2}{3}}}{f},
   \label{eq:m_sigma_square}
\ee
where $M_1$ is the \emph{material strength index} and $\psi_B^c$ and $\psi_B^y$ are the \emph{microscopic shape factors for failure in bending for buckling and yield}, respectively.

The microscopic shape factors $\psi_B^\delta$, $\psi_B^c$ and $\psi_B^y$ determine the optimal design. If these are smaller than one, there is no gain in introducing microstructure and the optimal volume fraction is $f=1$, i.e. a solid beam. If larger than one, mass is reduced by this factor by introducing microstructure. Assuming a one-term polynomial interpolation function, say $\tilde{E}_f\sim f^p$, microstructure is favourable with respect to elastic bending (\ref{eq:m_delta_square}) when $p<2$. Similarly, microstructure is favourable with respect to strength (both yield and buckling) if $p<3/2$ for those cases. This shows that a too high exponent (i.e. bad material performance at low densities) makes the solid beam preferable. Oppositely, low exponents, i.e. efficient materials, favour low volume fractions taking advantage of their optimal performance. Depending on exponents between stiffness and strength cases, intermediate volume fractions may become optimal. For the stretch-dominated microstructures considered here, the stiffness exponent is always one and hence microstructure is always favourable with regards to minimizing mass with a displacement constraint. For the strength case it is less simple.

For interpolation functions composed of two-term polynomials, both exponents should be below the numbers given above to favour porous material. If the lowest exponent is above, solid is preferred. If only the first exponent is below, porous material is preferred at least for lower volume fractions and solid may be preferred for higher volume fractions depending on the second multiplier.

The minimum mass beam satisfying both displacement and stress constraints, following \cite{Ash99}, is found by defining a coupling constant $C$ by equating (\ref{eq:m_delta_square}) and (\ref{eq:m_sigma_square}) and solving for $M_2$
\be
   M_2 = 384^{-\frac{1}{6}} \left( \frac{L^6}{V (\delta^*)^3} \right)^\frac{1}{6} M_1 = C M_1, \quad
         C = 384^{-\frac{1}{6}} \left( \frac{L^6}{V (\delta^*)^3} \right)^\frac{1}{6}.
         \label{eq:C_square}
\ee
By plotting material stiffness index $M_2$ versus material strength index $M_1$ for various microstructures and hierarchical levels, one can identify the optimal beam composition for given beam dimensions, displacement constraint and loading as defined by a line with slope $C$ in the corresponding Ashby plot.

The same study as above can be done for beams with fixed width and variable height or vise versa. For the former case, microscopic shape factors become $\psi_B^\delta = \tilde{E}_f^\frac{1}{3}/f$, $\psi_B^c = \tilde{\sigma}_{c,f}^{\frac{1}{2}}/f$ and $\psi_B^y = \tilde{\sigma}_{y,f}^{\frac{1}{2}}/f$, respectively (see Appendix  \ref{App:beam} for derivations). In this case, the simple exponents determining advantage of porous microstructure are 3 and 2 for the displacement and strength cases, respectively. This means that worse performing microstructures (higher exponents) than in the square cross section case are still advantageous compared to the solid beam. Naturally, mass of the optimal beam will hence also be lower than for the square cross section case. For the fixed height, variable width case microscopic shape factors become $\psi_B^\delta = \tilde{E}_f/f$, $\psi_B^c = \tilde{\sigma}_{c,f}/f$ and
$\psi_B^y = \tilde{\sigma}_{y,f}/f$, respectively (see Appendix  \ref{App:beam} for derivations). In this case, the simple exponents determining advantage of porous microstructure are 1 for both displacement and strength cases. This means that it is never advantageous to introduce porosity in the variable width case! Actually, it is a disadvantage because $\psi$ values always are below one for all microstructures (refer to $a_0$, $b_0$ and $c_0$ coefficients in Table \ref{tab:uniaxial}).

Considering simple tension/compression of a bar, microscopic shape factors are similar to the variable width problem, meaning that microstructure is never advantageous. Including stability for a square cross-sectioned column results in an added microscopic shape factor $\psi_B^{gc} = \tilde{E}_f^\frac{1}{2}/f$, which may or may not make microstructure favourable depending on slenderness ratio of the column.

\section{Example}

Based on above derivations we proceed to a practical example. As a test case we consider an Epoxy beam of length $L_0=1$m. As a baseline design, we give it cross-sectional dimensions $w_0=h_0=0.012$m. For the solid beam, this results in displacement $\delta_0=0.0235$m, mass $m_0=0.202$kg and maximum stress $\sigma_{max}=3.47$ MPa, i.e. well below the yield limit of Epoxy which is 72 MPa (c.f. Table \ref{tab:materialprops}). The displacement constraint is now selected as $\delta^*=\delta_0=0.0235$m for the remainder of the study.

\begin{table}[]
 \begin{center}
 \caption*{Material- properties and indices}
 \begin{tabular}{|p{2.7cm}|p{1.2cm}|p{1.2cm}|p{1.4cm}|p{1.cm} p{1.2cm}|p{1.cm} p{1.2cm}|p{1.cm} p{1.2cm}|}
  
  \hline
                & $E_0$ (GPa) & $\sigma_0$ (MPa) & $\rho_0$ (kg/m$^3$) & $\rho_0 / E_0^{\frac{1}{2}}$ ($\cdot 10^3$) & $\rho_0 / \sigma_0^{\frac{2}{3}}$ ($\cdot 10^3$) &  $\rho_0 / E_0^{\frac{1}{3}}$ & $\rho_0 / \sigma_0^{\frac{1}{2}}$ & $\rho_0 / E_0$ ($\cdot 10^6$) & $\rho_0 / \sigma_0$ ($\cdot 10^6$) \\
  \hline
  Pyrolytic Carbon \cite{CroBauVal20} & 62    & 2750 (7000) & 1400 & \textbf{5.6} & \textbf{0.71} (\textbf{0.38}) & \textbf{0.35} & \textbf{0.027}  (\textbf{0.017}) & \textbf{0.023} & \textbf{0.51} (\textbf{0.20}) \\
  \hline
  Steel       & \textbf{215} & 395 & 7800 & 16.8 & 14.5 & 1.30 & 0.39 & 0.036 & 19.7 \\
  \hline
  Epoxy         & 3.08 & 72 & 1400 & 25.2 & 8.10 & 0.96 & 0.17 & 0.45 & 19.5 \\
  \hline
  TPU         & 0.012 & 4.0 & \textbf{1190} & 344 & 47.2 & 5.20 & 0.60 & 99.2 & 298  \\
  \hline
 \end{tabular}
 \caption{Base material properties and associated material indices. Bold font indicates best property. Numbers in parentheses denote values that only work in compression.}
 \label{tab:materialprops}
 \end{center}
\end{table}

\begin{figure}[!b]
  \centering
  \includegraphics[width=12cm]{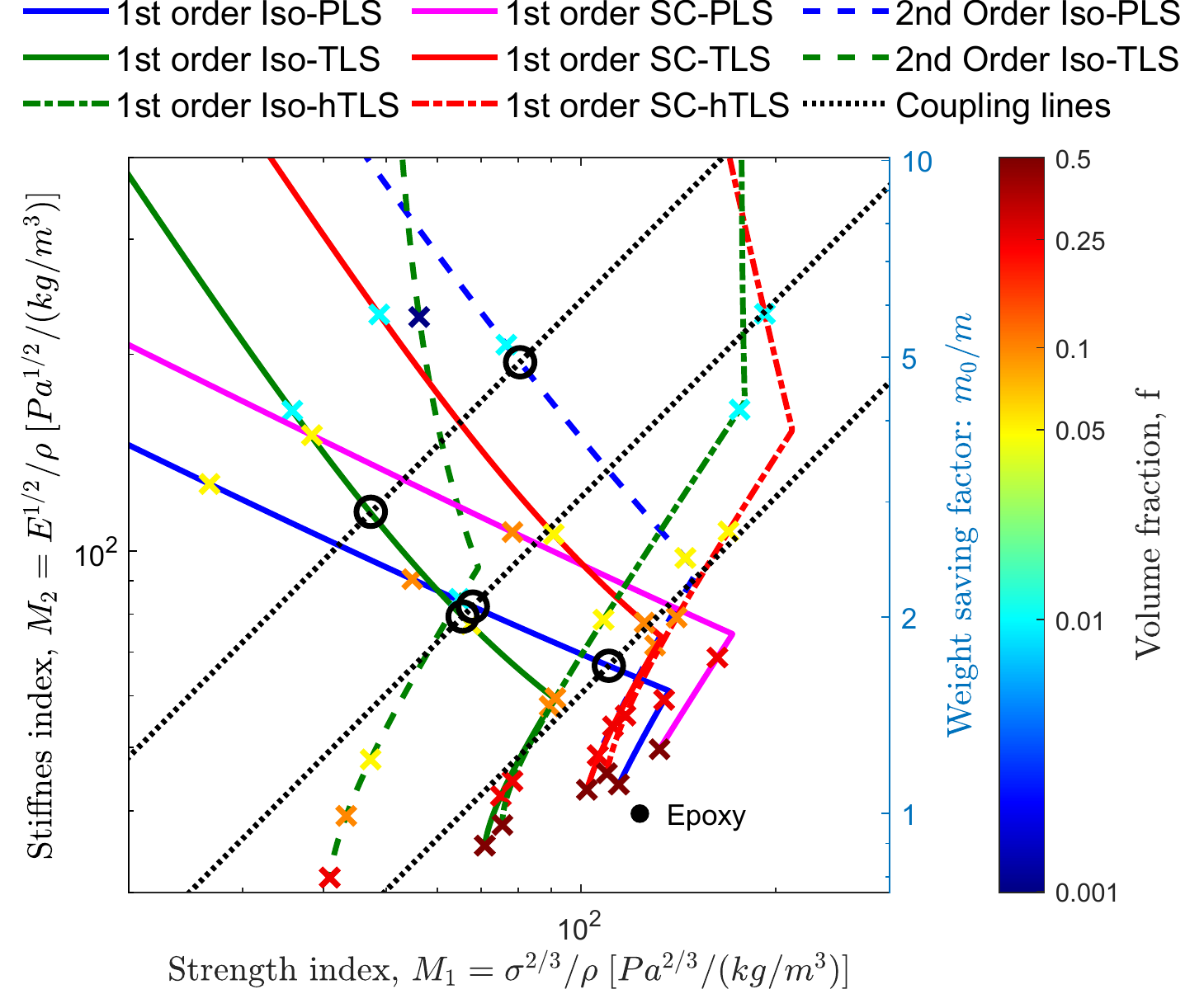}
  \caption{Ashby multi-objective chart for Iso-PLS (blue), Iso-TLS (green), SC-PLS (magenta) and SC-TLS (red) square cross-sectioned beams built from Epoxy. Solid colored curves indicate first order and dashed lines second order hierarchical microstructures and dash-dotted green and red curves indicate first order hTLS microstructures. The colored markers indicate volume fractions according to the color bar. Black dotted lines indicate coupling lines as discussed in the text and black circles indicate reference points for the specific beam examples discussed in the text and illustated in Figure \ref{fig:beams}. The right vertical axis indicates the weight saving factor compared to the solid Epoxy referencee beam.}
  \label{fig:ES-plots}
\end{figure}

Figure \ref{fig:ES-plots} shows the Ashby multi-objective plot for the Epoxy beam built from Iso-PLS (blue solid line), Iso-TLS (green solid line), SC-PLS (magenta solid line) and SC-TLS (red solid line) microstructures, respectively. Dashed lines indicate second order hierarchical versions and dash-dotted green and red curves denote hTLS. The graph includes three black dotted lines with the left most one corresponding to the coupling line for the reference beam (C=2.418 [Pa$^{-1/6}$]) obtained from inserting physical values in (\ref{eq:C_square}). The optimal beam is obtained for the microstructure curve that crosses the coupling line furthest to the north-east. The resulting weight saving can be read from the right y-axis. First considering isotropic microstructures, this happens for volume fraction $f=0.0778$ and $m=0.0781$kg (i.e. a mass saving factor of 2.6 with respect to the reference design) for the Iso-PLS case (solid blue curve) and $f=0.0212$ and $m=0.0699$kg (i.e. a 2.9 mass saving factor) for the Iso-TLS case (solid green curve). Hence here, the TLS provides the most efficient beam beating the plate lattice structure! Shortening the beam span to $L_0/2$ (with everything else the same as before), which corresponds to halving $C$  (center black dashed line), the resulting saving factors are 2.1 for Iso-PLS and 2.0 for Iso-TLS, respectively. Further shortening to $L_0/4$, corresponding to coupling factor $C/4$ (rightmost black dotted line), the resulting factors are 1.7 and 1.5 where the latter is determined by yield, as opposed to buckling as was the case in all previous cases. It is thereby demonstrated that depending on beam geometry, PLS or TLS may be preferred. PLS are desirable for the short span cases. Oppositely, TLS are desirable for long span beams. In both cases, it is buckling and not yield that controls dimensions. Only if making the beams even shorter than $L_0/4$ would yield become the controlling factor. Similar conclusions can be drawn for the simple cubic plate and truss lattice structures indicated by magenta and red solid lines in Figure \ref{fig:ES-plots}, respectively. Relaxing microstructure symmetry requirements from isotropy to cubic symmetry results in further weight savings and here the TLS (red) line again turns out as winner for the longer span cases. However, it should be remarked that the SC-TLS has very low shear stiffness and hence would fail for beam bending cases that have non-zero shear forces.

\begin{figure}[!t]
  \centering
  \includegraphics[width=16cm]{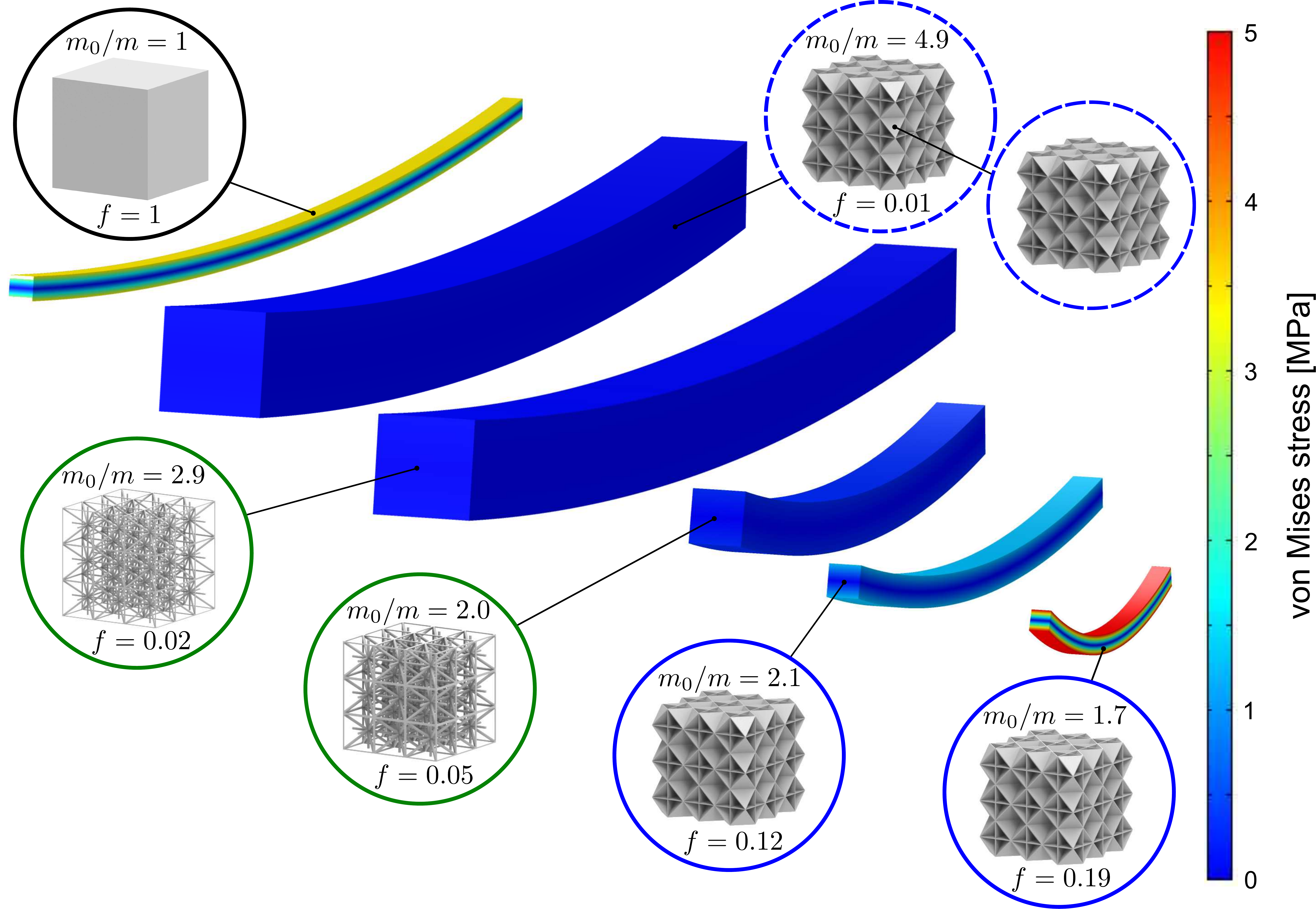}
  \caption{Plots of deformed beams and stresses for various cases of optimal isotropic microstructures. Compared to a solid Epoxy beam, large savings can be obtained by low density TLS structures for same length and higher density PLS for shorter lengths. Performance of each of the five cases are identified by black circles in Figure \ref{fig:ES-plots}.}
  \label{fig:beams}
\end{figure}

Above discussions seem to indicate a tie in the competition between truss and plate lattice structures. However, allowing for second order hierarchy changes the situation entirely. The dashed blue line in Figure \ref{fig:ES-plots}, corresponding to a second order hierarchical Iso-PLS structure, turns out to outperform all the other cases for both long and short beam lengths. The explanation for this is partly due to its volume fraction dependency on buckling which is a power of 2, corresponding to that of the simple TLS structures as discussed earlier. Partly, it comes from its $a_0$ factor (c.f. (\ref{eq:buck}) and Table \ref{tab:uniaxial}), which is much larger than for the TLS structures.

Finally, the hollow Iso-hTLS (dash-dotted green curve) beat all other microstructures for the long span beam with a weight saving factor exceeding 10. For shorter beam spans, the hollow simple cubic hTLS (dash-dotted red curve) provides the largest weight saving factor but as before for the SC-TLS, it is not applicable to general beam bending problems.

Sticking to simple isotropic microstructures (blue and green) and solid bars or plates (solid lines), the competition between TLS and PLS thus ends in a tie and depends on the structural application considered. Allowing for hierarchical structures, plate lattice structures turn out to be the optimal microstructures over the whole beam length range. Finally, if one is able to manufacture  the hTLS, with material inside crossings, those structures end up as the overall winners. Seen in the latter view, the competition is closed and turns out in favor of hollow truss structures hTLS, if one has the right manufacturing capabilities.

Figure \ref{fig:beams} gives a geometrical interpretation of above discussions, excluding the advanced hollow and hierarchical microstructures. By substituting solid Epoxy of the reference beam (left) with porous isotropic microstructure, long spans favor TLS and shorter spans favor PLS. Even more extreme weight savings of up to 4.9 can be obtained from a second order hierarchical PLS. Weight saving factors for each case are given in the table and corresponding performance points are indicated with black circles in Figure \ref{fig:ES-plots}.

\begin{figure}[!h]
  \centering
  \includegraphics[width=12cm]{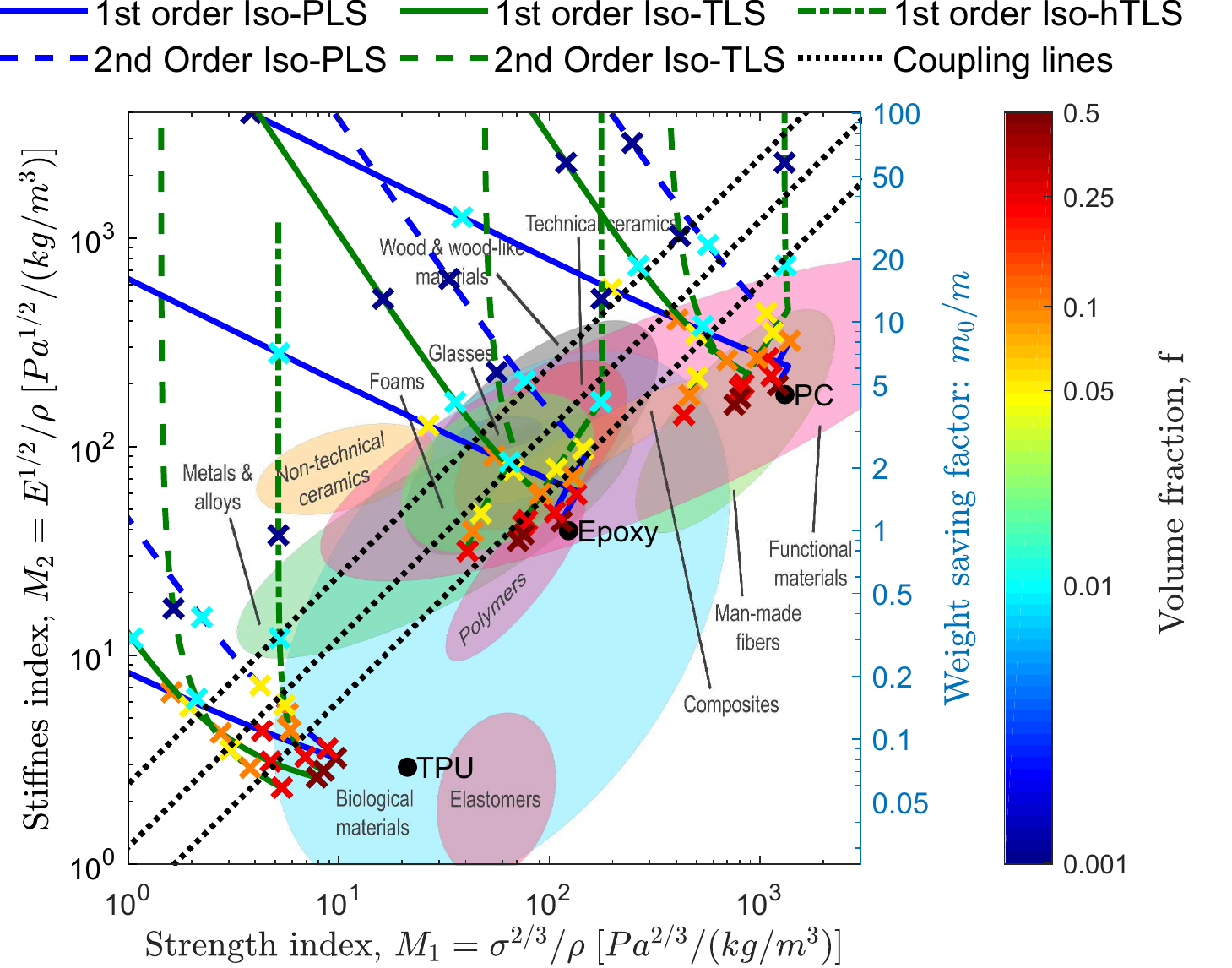}
  \caption{Ashby multi-objective chart for Iso-PLS (blue) and Iso-TLS (green) beams built from Epoxy, TPU and PC. Solid lines indicate first order and dashed lines indicate second order hierarchical microstructures. The dash-dotted green line refers to first order Iso-hTLS. The colored markers indicate volume fractions according to the color bar.}
  \label{fig:ES-plotsMultiSquare}
\end{figure}

Figure \ref{fig:ES-plotsMultiSquare} shows collected results for beams built from first and second order isotropic truss and plate microstructures as well as hTLS realized in Epoxy, TPU and Pyrolyzed Carbon (PC) on top of commonly encountered material property families. Remark here that the remarkable compressive yield strength of 7 GPa reported for nano-scale Iso-PLS PC microstructures by \cite{CroBauVal20} is not applicable here since beam bending inevitably involves both compression and tension. Hence, we use their reported yield strength value of $\sigma_0=2.75$ GPa for PC instead. Although transition points between TLS and PLS optimality vary slightly for different material choices due to varying ratios between base material stiffness and yield strengths, main observations from the simple Epoxy discussion still hold. However, it is remarkable that a weight saving factor approaching 20 compared to the solid Epoxy reference beam will be possible when technology allows to build large scale structures composed of Pyrolytic Carbon truss nano-lattice structures and even beyond 25 the day second order hierarchical plate nano-lattice structures can be realized in the same material. Having manufacturing capability to realize hTLS, one may obtain weight saving factors as high as 80 for a volume fraction of around 0.0005 in PC. On the other hand, TPU will never be a good material for beams and would potentially result in a beam more than 3 times heavier than the solid Epoxy beam for the same stiffness and strength requirements, even when using the highly efficient hTLS microarchitecture.

\section{Conclusions}

A "stiff and strong competition" has been going on within architected materials for decades, especially between open truss lattice structures and closed-walled plate lattice structures. By systematic studies of a number of high-performance candidate microstructures from the literature we conclude that there is no clear winner - at least not between simple truss and plate lattice structures. Depending on structural loading scenario one or the other type of microstructures may be preferred. For more complex hierarchical architectures, plate lattice structures beat truss lattice structures. However, if one is able to batch fabricate partially hollow truss lattice structures, these may beat all others for specific applications within beam bending.

Despite the extensive studies presented here, the search for ultimately stiff and strong microstructures is not over yet. One may consider using systematic topology optimization approaches that account for both microstructural stiffness as well as buckling response. Such a study was already performed in 2D \cite{ThoWanSig17} and resulted in intricate semi-hierarchical microstructures with much improved mechanical properties. Similarly, a systematic study in 3D may result in structures that are even stiffer and stronger than the "standard" geometries considered here. In this endeavour, one may consider including manufacturing constraints that reflect the manufacturing process at hand. An example is the topology optimization of multiscale graded structures as an extension of earlier pure stiffness design studies \cite{GroStuSig19}. Instead of the uniform beam structures discussed there, highly improved structures can be envisioned where, apart from varying local volume fractions, one could also locally identify the optimal microstructure depending on local stress state and hence always selected closed walled and optimally stiff PLS in tension regions and let the findings of the present study guide the choice of microstructure in the compression regions. The interpolation schemes provided here may hence pave the way for a new generation of multiscale design procedures that includes microstructural failure modes in the optimization process. We also expect that our findings can shed light on the appearances of open versus closed wall microstructures in natural structures. An obvious example being bone structures that often is open-celled and hence not optimal with respect to simple stiffness objectives in the low volume fraction limit. The reason for open-celled bone microstructure may be governed by length-scale effects, requirements to flow of nutrients, microstructural stability as studied here, or even by other driving goals.

\section*{Data Availability}
All numerical data for polynomial material interpolation coefficients (Table \ref{tab:uniaxial}) are available as SI.

\section*{Acknowledgements}
We acknowledge the financial support from the Villum Investigator Project InnoTop. We further acknowledge valuable discussions with Yiqiang Wang regarding add-up models and Niels Aage regarding finite element modelling.

\bibliography{Paper-V8}

\ifx \cedla \undefined \let \cedla = \c \fi\ifx \cyr \undefined \let \cyr =
  \relax \fi\ifx \cprime \undefined \def \cprime {$\mathsurround=0pt '$}\fi\ifx
  \prime \undefined \def \prime {'} \fi
\begin{thebibliography}{10}

\bibitem{Mil18}
Milton G (2018) Stiff competition.
\newblock {\em Nature} 564(7734):E1--E1.

\bibitem{HasSht63}
Hashin Z, Shtrikman S (1963) A variational approach to the theory of the
  elastic behaviour of multiphase materials.
\newblock {\em Journal of the Mechanics and Physics of Solids} 11:127--140.

\bibitem{FraMur86}
Francfort GA, Murat F (1986) Homogenization and optimal bounds in linear
  elasticity.
\newblock {\em Archive for Rational Mechanics and Analysis} 94(4):307--334.

\bibitem{Mil86}
Milton GW (1986) Modelling the properties of composites by laminates in {\em
  Homogenization and Effective Moduli of Materials and Media}, IMA, eds.{}
  Ericksen JL, Kinderlehrer D, Kohn R, Lions JL.
\newblock (Springer Verlag) Vol.{}~1, pp. 150--174.

\bibitem{Ave87-01}
Avellaneda M (1987) Optimal bounds and microgeometries for elastic two-phase
  composites.
\newblock {\em SIAM Journal on Applied Mathematics} 47(6):1216--1228.

\bibitem{Chr86}
Christensen R (1986) Mechanics of low density materials.
\newblock {\em Journal of the Mechanics and Physics of Solids} 34(6):563--578.

\bibitem{SigAagAnd16}
Sigmund O, Aage N, Andreassen E (2016) On the (non-)optimality of {M}ichell
  structures.
\newblock {\em Structural and Multidisciplinary Optimization} 54:361--372.

\bibitem{BerWadMcm17}
Berger J, Wadley H, McMeeking R (2017) Mechanical metamaterials at the
  theoretical limit of isotropic elastic stiffness.
\newblock {\em Nature} 543(7646):533--537.

\bibitem{TraSigGro18}
Tr\"{a}ff E, Sigmund O, Groen J (2019) Simple single-scale microstructures
  based on rank-3 optimal laminates.
\newblock {\em Structural and Multidisciplinary Optimization} 59(4):1021--1031.

\bibitem{WanGroSig18}
Wang Y, Groen J, Sigmund O (2019) Simple optimal lattice structures for
  arbitrary loadings.
\newblock {\em Extreme Mechanics Letters} 29:100447.

\bibitem{MezDasGre14}
Meza L, Das S, Greer J (2014) Strong, lightweight, and recoverable
  three-dimensional ceramic nanolattices.
\newblock {\em Science} 345:1322--6.

\bibitem{BauSchKra16}
Bauer J, Schroer A, Schwaiger R, Kraft O (2016) Approaching theoretical
  strength in glassy carbon nanolattices.
\newblock {\em Nature Materials} 15(4):438--443.

\bibitem{SchJacCar11}
Schaedler T, et~al. (2011) Ultralight metallic microlattices.
\newblock {\em Science} 334(6058):962--965.

\bibitem{ZhaVyaLi19}
Zhang X, Vyatskikh A, Gao H, Greer J, Li X (2019) Lightweight, flaw-tolerant,
  and ultrastrong nanoarchitected carbon.
\newblock {\em Proceedings of the National Academy of Sciences}
  116(14):6665--6672.

\bibitem{PorWidKoc20}
Portela C, et~al. (2020) Extreme mechanical resilience of self-assembled
  nanolabyrinthine materials.
\newblock {\em Proceedings of the National Academy of Sciences}
  117(11):5686--5693.

\bibitem{CroBauVal20}
Crook C, et~al. (2020) Plate-nanolattices at the theoretical limit of stiffness
  and strength.
\newblock {\em Nature communications} 11:1579.

\bibitem{TanDiaNoh18}
Tancogne-Dejean T, Diamantopoulou M, Gorji M, Bonatti C, Mohr D (2018) 3d
  plate-lattices: An emerging class of low-density metamaterial exhibiting
  optimal isotropic stiffness.
\newblock {\em Advanced Materials} 30(45):1803334.

\bibitem{TriNesSch06}
Triantafyllidis N, Nestorovi\'{c} MD, Schraad MW (2006) Failure surfaces for finitely
 strained two-phase periodic solids under general in-plane loading.
\newblock {\em Journal Applied Mechanics} 505-515.

\bibitem{MicLopCas07}
Michel JC, et~al. (2007) Microscopic and macroscopic instabilities in finitely
	strained porous elastomers.
\newblock {\em The Mechanics and Physics of Solids} 55(5):900-38.

\bibitem{LopCas07a}
Lopez-Pamies O, Casta\~{n}eda PP (2007) Homogenization-based constitutive models for 
	porous elastomers and implications for macroscopic instabilities: I-Analysis. 
\newblock {\em Journal of the Mechanics and Physics of Solids} 55(8):1677-1701.

\bibitem{LopCas07b}
Lopez-Pamies O, Casta\~{n}eda PP (2007) Homogenization-based constitutive models for 
	porous elastomers and implications for macroscopic instabilities: II-Results. 
\newblock {\em Journal of the Mechanics and Physics of Solids} 55(8):1702-1728.

\bibitem{Bit97}
Bitzer, TN (1997) Honeycomb technology: materials, design, manufacturing, 
	applications and testing.
\newblock {\em Springer Science \& Business Media}.

\bibitem{GroStuSig19}
Groen J, Stutz F, Aage N, B{\ae}rentzen J, Sigmund O (2020) De-homogenization
  of optimal multi-scale 3d topologies.
\newblock {\em Computer Methods in Applied Mechanics and Engineering}
  364:112979.

\bibitem{Cas91}
Casta\~{n}eda PP (1991) The effective mechanical properties of nonlinear isotropic composites.
\newblock {\em Journal of the Mechanics and Physics of Solids} 39(1):45-71.

\bibitem{Suq93}
Suquet P (1993) Overall potentials and extremal surfaces of power law or
  ideally plastic composites.
\newblock {\em Journal of the Mechanics and Physics of Solids} 41(6):981 --
  1002.

\bibitem{GibAsh97}
Gibson LJ, Ashby MF (1997) {\em Cellular Solids, Structure and Properties,
  Second Edition}.
\newblock (Cambridge University Press, Cambridge, England).

\bibitem{GurDur14}
Gurtner G, Durand  (2014) Stiffest elastic networks.
\newblock {\em Proceedings of the Royal Society A: Mathematical, Physical and
  Engineering Sciences} 470(2164):20130611.

\bibitem{ZheSmiSpa16}
Zheng X, et~al. (2016) Multiscale metallic metamaterials.
\newblock {\em Nature Materials} 15:1100-1106.

\bibitem{WeaAsh97}
Weaver P, Ashby M (1997) Material limits for shape efficiency.
\newblock {\em Progress in materials science} 41(1-2):61--128.

\bibitem{TriSch93}
Triantafyllidis N, Schnaidt WC (1993) Comparison of microscopic and macroscopic
  instabilities in a class of two-dimensional periodic composites.
\newblock {\em Journal of the Mechanics and Physics of Solids} 41(9):1533--65.

\bibitem{NevSigBen01}
Neves MM, Sigmund O, Bends{\o}e MP (2002) Topology optimization of periodic
  microstructures with a penalization of highly localized buckling modes.
\newblock {\em International Journal for Numerical Methods in Engineering}
  54(6):809--834.

\bibitem{ThoWanSig17}
Thomsen C, Wang F, Sigmund O (2018) Buckling strength topology optimization of
  {2D} periodic materials based on linearized bifurcation analysis.
\newblock {\em Computer Methods in Applied Mechanics and Engineering}
  339:115--136.

\bibitem{HagPapVaz14}
Haghpanah B, Papadopoulos J, Mousanezhad D, Nayeb-Hashemi H, Vaziri A (2014)
  Buckling of regular, chiral and hierarchical honeycombs under a general
  macroscopic stress state.
\newblock {\em Proceedings of the Royal Society A: Mathematical, Physical and
  Engineering Sciences} 470(2167):20130856.

\bibitem{BluSigPou20}
Bluhm G, Sigmund O, Wang F, Poulios K (2020) Nonlinear compressive stability of
  hyperelastic 2d lattices at finite volume fractions.
\newblock {\em Journal of the Mechanics and Physics of Solids} 137:103851.

\bibitem{Ash99}
Ashby M (1999) {\em Materials Selection in Mechanical Design (Second Edition)}.
\newblock (Butterworth-Heinemann, Oxford), Fourth edition edition.

\bibitem{Lak93}
Lakes R (1993) Materials with structural hierarchy.
\newblock {\em Nature} 361:511--515.

\bibitem{BouKoh08}
Bourdin B, Kohn RV (2008) Optimization of structural topology in the
  high-porosity regime.
\newblock {\em Journal of the Mechanics and Physics of Solids}
  56(3):1043--1064.

\bibitem{OhnOkuNii04}
Ohno N, Okumura D, Niikawa T (2004) Long-wave buckling of elastic square
  honeycombs subject to in-plane biaxial compression.
\newblock {\em International journal of mechanical sciences} 46(11):1697--1713.

\bibitem{Bud02}
Budynas RG, Young WC, Sadegh A (2002) Roark's formulas for stress and strain.
\newblock {\em McGrwhill Publication}.

\bibitem{Sig94}
Sigmund O (1994) Materials with prescribed constitutive parameters: an inverse
  homogenization problem.
\newblock {\em International Journal of Solids and Structures}
  31(17):2313--2329.

\end{thebibliography}
\appendix
\section{Hashin-Shtrikman bounds}\label{App:bounds}

The upper Hashin-Shtrikman bounds \cite{HasSht63} on Young's modulus for isotropic and cubic symmetric microstructures are
\be
   E^u_{Iso} = \frac{2 f (7 - 5\nu)}{(15 \nu_0^2 + 2 \nu_0 - 13) f - 15 \nu_0^2 - 12 \nu_0 + 27} \ {\rm and} \
   E^u_{Cubic} = \frac{2 f (2-\nu_0)}{(\nu_0^2+\nu_0-2) f - 3 \nu_0^2 - 3 \nu_0+6}.
\ee
Inserting $\nu_0=1/3$ in these expressions results in the simplified versions given in (\ref{eq:HS}).

\section{Discussion on Interpolations for hierarchical microstructures}\label{App:hierarchical}

This appendix, first repeats (and corrects a typo in) Lakes derivations for material properties of hierarchical microstructures based on the simple one-term material interpolation scheme. Then it extends the derivations to the recommended two-term interpolation scheme and a fully general interpolation scheme.

\subsection{One-term interpolation derivation}
Lakes \cite{Lak93} based his derivations on the one-term interpolation scheme for Young's modulus
\begin{equation}
E_1= a_0 f^{m_0} E_0
\end{equation}
where $m_0$ is the general exponent governing the density response. In this paper we use $m_0=1$ for stretch-dominated microstructures but in this appendix we keep $m_0$ as an open parameter for generality.
The Young's modulus of an $n$'th order hierarchical microstructure is found as
\begin{equation}
E_n= a^n_0 (\rho_n/\rho_{n-1})^{m_0} \ldots (\rho_1/\rho_{0})^{m_0}  E_0 = a^n_0 (\rho_n/\rho_0)^{m_0} E_0 =a^n_0 f^{m_0} E_0
\end{equation}

The buckling strength of a first order hierarchical microstructure is interpolated by
\begin{equation}
\sigma_{c,1}=b_0f^{n_0}E_0
\end{equation}
Following the same idea, the strength of an  $n$'th order hierarchical microstructure is found as
\begin{equation}
\sigma_{c,n}=b_0\left(\frac{\rho_n}{\rho_{n-1}}\right)^{n_0} E_{n-1}
\end{equation}

Assuming a self-similar hierarchy, i.e. the same volume fraction and microstructure at each level, one obtains $\frac{\rho_n}{\rho_{n-1}}=f^{\frac{1}{n}}$. The strength of an $n$'th order hierarchical microstructure is thus
\begin{equation}
\sigma_{c,n}=b_0\left(\frac{\rho_n}{\rho_{n-1}}\right)^{n_0}E_{n-1}=b_0f^{\frac{n_0}{n}} a^{n-1}_0 f^{\frac{{m_0}(n-1)}{n}}  E_0 =b_0a^{n-1}_0f^{m_0+\frac{n_0-m_0}{n}}E_0
\end{equation}
Note here that Lakes' paper used $n$ instead of $n-1$ for the exponent on $a_0$, which is a typo.

The yield strength of a first order hierarchical microstructure is interpolated by
\begin{equation}
\sigma_{y,1}=c_0f^{p_0}\sigma_0
\end{equation}

For an $n$'th order microstructure, the relation between the yield strength in the $n$-level (global level) and the $n-1$ level becomes
\begin{equation}
\sigma_{y,n}=c_0\left(\frac{\rho_n}{\rho_{n-1}}\right)^{p_0}\sigma_{y,n-1}
\end{equation}
With this, the yield strength for the $n$'th order microstructure is
\begin{align}
\sigma_{y,n}&=c_0\left(\frac{\rho_n}{\rho_{n-1}}\right)^{p_0}\sigma_{n-1}  =c_0\left(\frac{\rho_n}{\rho_{n-1}}\right)^{p_0} c_0\left(\frac{\rho_{n-1}}{\rho_{n-2}}\right)^{p_0}  \ldots \ c_0  \left(\frac{\rho_{1}}{\rho_{0}}\right)^{p_0} \sigma_0
=c^{n}_0f^{p_0}  \sigma_0
\end{align}

\subsection{Two-term interpolation}
Using the proposed two-term material interpolation scheme, the  Young's modulus follows
\begin{equation}
E_1= \left( a_0 f^{m_0} +a_1f^{m_0+1}\right)  E_0
\end{equation}
Assuming the same volume fraction at each level, the Young's modulus of an $n$'th order hierarchical microstructure is interpolated by

\begin{equation}
E_n=\left(   a_0 \left(\frac{\rho_n}{\rho_{n-1}}\right)^{m_0}+  a_1\left(\frac{\rho_n}{\rho_{n-1}}\right)^{m_0+1}\right)  E_{n-1}=\left(  a_0+a_1f^{\frac{1}{n}}\right) ^n f^{m_0}  E_0
\end{equation}

The buckling strength of a  first order hierarchical microstructure is interpolated by
\begin{equation}
\sigma_{c,1}=\left( b_0f^{n_0}  +b_1f^{n_0+1}\right) E_0
\end{equation}
The buckling strength of an $n$'th order hierarchical microstructure  is
\begin{align}
\sigma_{c,n}&=\left( b_0\left(\frac{\rho_n}{\rho_{n-1}}\right)^{n_0} +b_1\left(\frac{\rho_n}{\rho_{n-1}}\right)^{n_0+1}\right) E_{n-1} \nonumber\\
&=\left(b_0+b_1 f^{\frac{1}{n}} \right) f^{\frac{n_0}{n}} (a_0+a_1f^{\frac{1}{n}})^{n-1} f^{\frac{{m_0}(n-1)}{n}}   E_0 \nonumber\\
&=\left(b_0+b_1 f^{\frac{1}{n}} \right) (a_0+a_1f^{\frac{1}{n}})^{n-1} f^{m0+\frac{n0-m0}{n}}E_0
\end{align}

The yield strength for a first order microstructure follows:
\begin{equation}
\sigma_{y,1}=\left( c_0 f^{p_0}  + c_1 f^{p_0+1}\right)  \sigma_0
\end{equation}
For an $n$'th order microstructure, the relation between the yield strengths in the $n$-level (global level) and the $n-1$ level is as following
\begin{equation}
\sigma_{y,n}= \left( c_0\left(\frac{\rho_n}{\rho_{n-1}}\right)^{p_0}  + c_1 \left(\frac{\rho_n}{\rho_{n-1}}\right)^{p_0+1}\right) \sigma_{y,n-1}=\left( c_0 f ^{\frac{p_0}{n}} + c_1  f ^{\frac{p_0+1}{n}}\right) \sigma_{y,n-1}
\end{equation}
Hence the yield strength for the $n$'th order microstructure is
\begin{align}
\sigma_{y,n}&=\left( c_0 f ^{\frac{p_0}{n}} + c_1  f ^{\frac{p_0+1}{n}}\right) \sigma_{y,n-1}
=\left( c_0 f ^{\frac{p_0}{n}} + c_1  f ^{\frac{p_0+1}{n}}\right)^n  \sigma_0
=\left(c_0+c_1 f^{\frac{1}{n}} \right)^nf^{p_0}\sigma_0
\end{align}

\subsection{General form}
Finally, and for completeness, we consider a fully general form of interpolation function
\begin{equation}
   E_1= \tilde{E}(f) E_0, \quad
   \sigma_{c,1}=\tilde{\sigma}_c(f)  E_0, \quad
   \sigma_{y,1}=\tilde{\sigma}_y(f) \sigma_0
\end{equation}
where $\tilde{E}(f)$, $\tilde{\sigma}_c(f)$ and $\tilde{\sigma}_y(f)$ are functions mapping the volume fraction to the relative material property for stiffness, buckling and yield strength, respectively. For example $\tilde{E}(f)=a_0 f^{m_0}$ for the one-term polynomial interpolation functions discussed above.

Again assuming a self-similar hierarchical structure (same microstructure and volume fraction at each level), the Young's modulus of the $n$'th order hierarchical microstructure using the general form  is
\begin{equation}
E_n = \left( \tilde{E}\left( f^{\frac{1}{n}} \right)\right)^n E_0
\end{equation}
The buckling strength of the $n$'th order hierarchical microstructure  is
\begin{align}
\sigma_{c,n}=\tilde{\sigma}_c\left(f^{\frac{1}{n}} \right) \left( \tilde{E}\left( f^{\frac{1}{n}} \right)\right)^{n-1} E_0
\end{align}
and the yield strength is
\begin{align}
\sigma_{y,n}= \left( \tilde{\sigma}_y\left(f^{\frac{1}{n}}  \right)\right)^n \sigma_0
\end{align}

\section{Analytical buckling studies}\label{App:hollow}

This appendix summarizes analytical expressions for buckling response of truss lattice structures. The same expressions are used to come up with shell thickness to strut cross-sectional dimensions ratio that prevents wall-buckling in the suggest hollow truss lattice structures.

\subsection{SC-TLS}
Based on the simple add up model~\cite{Chr86,BouKoh08},  the volume fraction of the SC-TLS is calculated as
\begin{equation} \label{eq:SCTLSVol}
f= 3A/l^2
\end{equation}
with $A$ being the cross-sectional area and $l$ being the microstructure size.
The effective Young's modulus is
\begin{equation} \label{eq:EH}
{E}=\frac{1}{3}fE_0
\end{equation}
The maximum von Mises stress under uni-axial compression of  $\boldsymbol{\sigma}^0=\left[\sigma_1,0,0,0,0,0 \right]^T$ is simply
\begin{equation}
{\sigma_{vm}}= \frac{3 \sigma_1  }{f}
\end{equation}
Hence, the corresponding yield strength is
\begin{equation}
{\sigma}_{y}= \frac{1}{3}f \sigma_0
\end{equation}
Following \cite{OhnOkuNii04},  the critical buckling strength  of the SC-TLS  under  uni-axial compression  due to a global shear failure is
\begin{equation} \label{eq:buckh}
{\sigma}_c= \frac{6E_0 I_b}{l^4}
\end{equation}
where $I_b$ is the second moment of area.  If the bars are  solid with a square cross-section, i.e., $I_b=A^2/12$,   the corresponding buckling strength of the  SC-TLS  is
\begin{equation} \label{eq:buckh}
{\sigma}_c= \frac{6}{108}  f^2  E_0 \approx 0.0556   f^2  E_0
\end{equation}

If the bars are hollow with a thin box cross-section with the dimension of $h$ and a uniform thickness of $t$, and ignoring higher order contributions of $t$, the corresponding cross-sectional area and second moment of area are calculated as $A=4ht$ and $I_b= {2 h^3t }/{3}$. The buckling strength of SC-hTLS due to the global shear failure, ${\sigma}_b$, and the buckling strength due to local wall buckling failure~\cite{WeaAsh97}, ${\sigma}_l$, are expressed as
\begin{align}
{\sigma}_b =&\frac{6 E_0 I_b}{l^4}=  \frac{  h^2  E_0 f}{3 l^2} \label{Eq:glBox}\\
   {\sigma}_l=& \frac{f}{3}   \frac{3.6 E_0t^2}{h^2}   =  \frac{E_0 f^3 l^4}{120h^4} \label{Eq:localBox}
\end{align}
The critical buckling strength of SC-hTLS is determined  by
\begin{equation}
{\sigma}_{c}=\min\left( {\sigma}_b,{\sigma}_l\right)
\end{equation}
The optimum is obtained when ${\sigma}_b={\sigma}_l$~\cite{WeaAsh97}, where the characteristic dimensions  of the optimal thin box-section are obtained by solving ${\sigma}_b={\sigma}_l$, written as
\begin{equation}\label{Boxsize}
h=\frac{\sqrt[3]{f}}{\sqrt[6]{40} } l,\qquad t=\frac{\sqrt[6]{40} \sqrt[3]{f^2}}{12 } l
\end{equation}
{The corresponding second moment of area is written as
 \begin{equation}\label{eq:optMoment}
 I_b=\frac{2 h^3t }{3}= \frac{1 }{36\sqrt[3]{5}} f^{\frac{5}{3}} l^4
 \end{equation}
}
By inserting $h$ in Eq. \eqref{Boxsize} into Eq.~\eqref{Eq:glBox}, one obtains
 \begin{equation}\label{eq:optBox}
 {\sigma}_{c}={\sigma}_b= \frac{1}{6\sqrt[3]{5}} f^{\frac{5}{3}}E_0  \approx  0.0975 f^{\frac{5}{3}}E_0
 \end{equation}

If the bars are hollow with a thick box cross-section, the corresponding area and second moment of area are $A=h^2_{out}-h^2_{in}$ and $I_b=\left(h^4_{out}-h^4_{in}\right) /12$, where $h_{out}$ and $h_{in}$ are the outer and inner dimensions, respectively. The corresponding buckling strength is determined by the global shear failure, given as
\begin{equation} \label{eq:buckh}
{\sigma}_{b,c}= \frac{ 6 E_0 I_b}{l^4}=\frac{   E_0 \left(h^4_{out}-h^4_{in}\right)  }{2l^4}=\frac{  \left(h^2_{out}+h^2_{in}\right)  }{6l^2}f E_0
\end{equation}
Now, we select the dimensions of the thick box cross-section  for SC-hTLS with a higher volume fraction such that its second moment of area follows the same function of $f$ as the thin wall box cross-section, i.e., $I_b=\frac{1 }{36\sqrt[3]{5}} f^{\frac{5}{3}} l^4 $ (see Eq.~\eqref{eq:optMoment}).   Together with the volume fraction equation, $f= 3A/l^2=3\left( h^2_{out}-h^2_{in}\right)/l^2 $, the dimensions of the thick box section are obtained as
 \begin{equation}
 \left\lbrace \begin{array}{ll}
3\left(  h^2_{out}-h^2_{in}\right)  &= fl^2     \\
   \left( h^4_{out}-h^4_{in}\right)/12 &  =  \frac{1 }{36\sqrt[3]{5}} f^{\frac{5}{3}}  l^4
 \end{array} \Rightarrow \left\lbrace    \begin{array}{ll}
 h_{out}&=  \sqrt{ \frac{f}{6}+\frac{ \sqrt[3]{f^2}}{2\sqrt[3]{5}} } l  \\
 h_{in} & = \sqrt{ \frac{ \sqrt[3]{f^2}}{2\sqrt[3]{5}} -\frac{f}{6} }l
 \end{array} \right. \right.
 \end{equation}
 The dimensions of the thick box section in SC-hTLS are shown as in Figure~\ref{Int_SC_TLS}.

\subsection{Iso-TLS}
Based on the simple add up model~\cite{Chr86,BouKoh08},  the volume fraction of the Iso-TLS is calculated by
\begin{equation}
f= 15 A/l^2
\end{equation}
where  $A$ is the cross-sectional area of the SC bars.
The effective Young's modulus is written as
\begin{equation} \label{eq:EH}
{E}=\frac{1}{6}fE_0
\end{equation}
The maximum von Mises stress under uni-axial compression  of  $\boldsymbol{\sigma}^0=\left[\sigma_1,0,0,0,0,0 \right]^T$ is stated
\begin{equation}\label{eq:ISOTLSVol}
{\sigma_{vm}}= \frac{6 \sigma_1  }{f}
\end{equation}
Hence, the corresponding yield strength is
\begin{equation}
{\sigma}_{y}= \frac{1}{6}f \sigma_0
\end{equation}
The critical buckling for the Iso-TLS under uni-axial compression is dominated by buckling of the SC bars with clamped-clamped boundaries. The critical buckling strength of the Iso-TLS is written as
\begin{equation} \label{eq:buckc}
{\sigma}_c=\frac{f}{6} {\sigma}_l=\frac{10 \pi^2E_0I_{sc}}{  l^4  }
\end{equation}
where $ {\sigma}_l$ is the buckling strength of the SC bars, and  $I_{sc}$ is the second moment of area of the SC bars. If all the bars are solid{ with circular cross-section, i.e., $I_{sc}=A^2/\left(4 \pi \right) $,} the corresponding buckling strength is
\begin{equation}
{\sigma}_c= \frac{\pi }{90}E_0f^2  \approx 0.0349 f^2 E_0
\end{equation}
If all the bars are thin tubes, the radius and thickness of  the SC bars  are $r$ and $t$, respectively. {Ignoring higher order contributions of t, the} corresponding cross-sectional area and second moment of area are $A=2\pi r t$ and $I_t= \pi r^3 t$. The buckling strength of the Iso-hTLS due to global buckling failure and the buckling strength due to local wall buckling failure~\cite{Bud02} are written as
\begin{align}
{\sigma}_h&=\frac{10 \pi^2E_0I_t}{  l^4 }= \frac{\pi^2   r^2 f }{3 l^2}  E_0, \label{Eq:gltube} \\
 {\sigma}_l&= \frac{f}{6} \left[ \frac{\alpha}{\sqrt{3\left( 1-\nu_0^2\right) }} E_0\frac{t}{r}\right] E_0  = \frac
 {\alpha f^2 l^2}{ 180\pi \sqrt{3\left( 1-\nu_0^2\right)} r^2}  E_0  \label{eq:thintube}
\end{align}
where $\nu_0=1/3$ is the Poisson's ratio of the base material.  Different from the local wall buckling of bars with a  thin box cross-section in
Eq.~\eqref{Eq:localBox}, the critical stress of ä thin tube  due to the  local wall buckling failure actually developed is usually only 40-60\% of the theoretical value~\cite{Bud02}. Hence a knock down  factor, $\alpha$,  is contained in Eq.~\eqref{eq:thintube}. Based on numerical buckling simulations, the knock down factor is here chosen as $\alpha=0.45$ for the Iso-hTLS. The critical buckling strength of the Iso-hTLS is thus
\begin{equation}
{\sigma}_{c}=\min\left( {\sigma}_h,{\sigma}_l\right)
\end{equation}
The optimal buckling strength is obtained by ${\sigma}_h={\sigma}_l$~\cite{WeaAsh97}, where the characteristic dimensions  of the optimal thin tube are obtained by solving ${\sigma}_h={\sigma}_l$, written as
\begin{equation}\label{tubesize}
r=\sqrt[4]{\frac{\alpha f }{100 \sqrt{ 1.08\left( 1-\nu_0^2\right) }\pi^3}} l \approx 0.110 \sqrt[4]{f}l , \qquad  t=\frac{  \sqrt[8]{ 1.08\left( 1-\nu_0^2\right) f^6} }{3 \sqrt[4]{100\alpha \pi }}  l\approx 0.096 \sqrt[4]{f^3}l
\end{equation}
The corresponding second moment of area is written as
{\begin{equation}\label{eq:opttubeMoment}
I_t= \pi r^3t= \frac{\sqrt{\alpha}}{300\sqrt{\pi^3} \sqrt[4]{ 1.08\left( 1-\nu_0^2\right)}}f^{1.5} l^4 \approx 0.00040  f^{1.5}  l^4
\end{equation}}
By inserting $r$ in Eq. \eqref{tubesize} into Eq.~\eqref{Eq:gltube}, we obtain
\begin{equation}\label{eq:optTube}
{\sigma}_c ={\sigma}_h =\frac{\sqrt{\alpha\pi}}{30 \sqrt[4]{1.08\left( 1-\nu_0^2\right) }}f^{1.5}E_0 \approx  0.040 f^{1.5}E_0
\end{equation}
If all the bars are thick tubes,  the  cross-sectional area of the SC bars is $A=\pi\left( r^2_{out}-r^2_{in}\right) $, where $r_{out}$ and $r_{in}$ are the outer and inner radii, respectively. The corresponding second moment of area is $ I_t= {\pi}/{4}  \left(r^4_{out}- r^4_{in}  \right)$.  The critical buckling strength of the  Iso-hTLS consisting of thick tubes is given as
\begin{equation} \label{eq:buckhol}
{\sigma}_{h,c}= \frac{10 \pi^2E_0I_{t}}{  l^4  } =   \frac{ \pi^2\left( r^2_{out}+r^2_{in}\right) }{6 l^2}  fE_0
\end{equation}

{As in the SC-hTLS case,  we select  the dimensions of the thick tube in the Iso-hTLS with a higher volume fraction such that its second moment  of area follows the same function of $f$ as the thin tube, i.e., $I_t=0.00040  f^{1.5}  l^4 $ (see Eq.~\eqref{eq:opttubeMoment}).} Together with the volume fraction equation, $f= 15 A/l^2=15 \pi\left( r^2_{out}-r^2_{in}\right)/l^2 $, the dimensions of the thick tube section are obtained as,
\begin{equation}
\left\lbrace \begin{array}{ll}
15\pi \left( r^2_{out}-r^2_{in}\right)   &=  {fl^2} \\
 \pi \left(  r^4_{out}-r^4_{in}\right)/4 &=  0.00040f^{1.5}l^4
\end{array} \Rightarrow \left\lbrace    \begin{array}{ll}
r_{out}&=  \sqrt{ 0.012 \sqrt{f}+\frac{f}{30\pi }} l  \\
r_{in} & =  \sqrt{ 0.012 \sqrt{f}  -\frac{f}{30\pi }} l
\end{array} \right. \right.
\end{equation}

\begin{figure}[!t]
	\centering
	\includegraphics[width=0.8\textwidth]{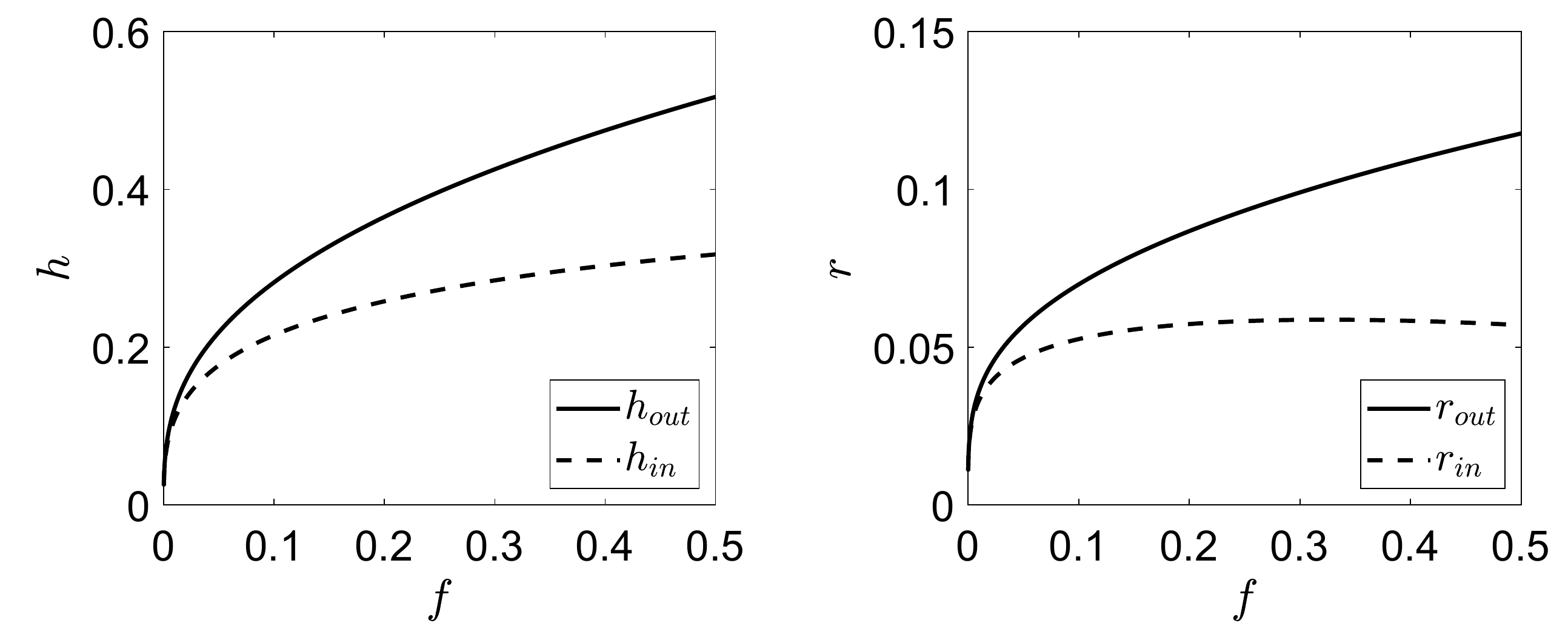}
	\caption{Hollow truss geometry. Dimensions of the box section in the SC-hTLS (left) and the tube in the Iso-hTLS (right).}\label{Int_SC_TLS}
\end{figure}
The dimensions of the thick tube  section in the Iso-hTLS  are shown as in Figure~\ref{Int_SC_TLS}.

\section{Numerical calculations}\label{App:numerics}
All the FE calculations in the study are conducted using COMSOL. The effective material properties are calculated using the homogenization method~\cite{Sig94}. The effective elasticity matrix, $\overline{\mathbf{D}}$,  is calculated as
\begin{align}
\overline{D}_{ij} = \frac{1}{|Y|} \sum_{e=1}^N   \int_{Y^e} \left(\tilde{\boldsymbol{\varepsilon}}_{i} - \mathbf{B}^e\boldsymbol{\chi}^e_{i} \right)^T  \mathbf{D}^e \left(\tilde{\boldsymbol{\varepsilon}}_{j} - \mathbf{B}^e\boldsymbol{\chi}^e_{j} \right) dY,
\end{align}
where the sum represents a finite element assembly operation over $N$ elements,  $\mathbf{B}^e$ is the strain-displacement matrix of element $e$, $\mathbf{D}^e$ is the elasticity matrix of the material in element $e$, which is the elasticity matrix of the base material in this study, i.e., $\mathbf{D}^e=\mathbf{D}^0$,   $\tilde{\varepsilon}_j=\delta_{ij}$ denotes the 6 independent unit strain fields, and $\boldsymbol{\chi}_{j}$ is the perturbation field induced by the $j$'th unit strain field under  periodic boundary conditions, solved by
\begin{align}
\mathbf{K}_0 \boldsymbol{\chi}_{j} = \mathbf{f}_{j}, \quad j=1,2,3,4,5,6.
\label{eq:chi_disc}
\end{align}
The initial stiffness matrix,  $\mathbf{K}_0$ and the equivalent load vectors $\mathbf{f}_{j}$ are given by:
\begin{align}
\mathbf{K}_0 = \sum_{e=1}^N \int_{Y^e} \left( \mathbf{B}^e\right) ^T \mathbf{D}^e  \mathbf{B}^e dY,
\quad
\mathbf{f}_{j} = \sum_{e=1}^N \int_{Y^e} \left( \mathbf{B}^e\right)^T \mathbf{D}^e \tilde{\boldsymbol{\varepsilon}}_{j} dY,
\label{eq:F_equiv}
\end{align}
The effective Young's modulus is calculated using the effective elasticity matrix $\overline{\mathbf{D}}$.

For a prescribed  macroscopic stress state, $ \boldsymbol{\sigma}^0$, the effective elasticity matrix is used to transform the macroscopic stress $\boldsymbol{\sigma}^0$ to macroscopic strain $\boldsymbol{\varepsilon}^0$ , then to the prestress of element $e$,   $\boldsymbol{\sigma}_0^e$, given as
	\begin{align}
	\boldsymbol{\sigma}^e_0 = \boldsymbol{D}^e  \boldsymbol{\varepsilon}^0  =\boldsymbol{D}^e \overline{\boldsymbol{D}}^{-1}\boldsymbol{\sigma}^0.\label{eq:chi_disc}
	\end{align}
The stress distribution in the microstructure under the macroscopic stress state  $ \boldsymbol{\sigma}^0$ is calculated by employing the prestress state, $\boldsymbol{\sigma}^e_0$  with periodic boundary conditions in COMSOL.

Based on the stress distribution in the microstructure, the yield strength under an uni-axial compression  of  $\boldsymbol{\sigma}^0=\left[\sigma_1,0,0,0,0,0 \right]^T$ is defined as
\begin{equation}
\sigma_y=\frac{\sigma_0}{\sigma_{vm}} \sigma_1
\end{equation}
where $\sigma_0$ is the yield stress of the base material, and $\sigma_{vm}$ is  the maximum  von Mises stress in the microstructure, which is here approximated by  the von Mises stress  at the middle of the plate/truss members. Thus, it does not account for stress concentrations in the microstructure.

\begin{figure}[!b]
	\begin{center}
		\includegraphics[height=0.3\textwidth]{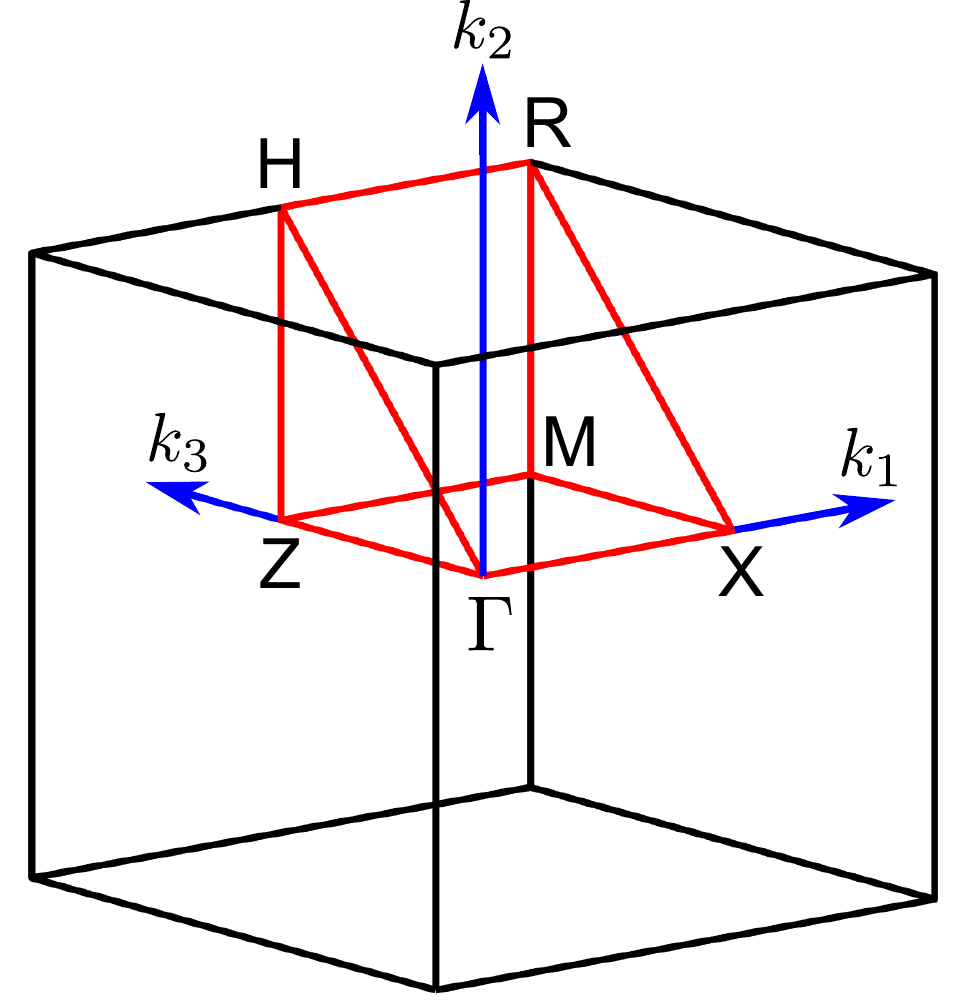}
	\end{center}
	\caption{Illustration of the irreducible Brillouin zone for uni-axial stress for  microstructures with cubic symmetry. Irreducible Brillouin zone: the region enclosed by the red lines. }  \label{Fig:IRZ}
\end{figure}

Subsequently, linear buckling analysis is performed  to evaluate the material buckling strength. Both short- and long-wavelength instabilities are captured by employing Floquet-Bloch boundary conditions in the linear buckling analysis~\cite{TriSch93,NevSigBen01,ThoWanSig17}. The material buckling strength of a unit cell is calculated by
\begin{align}
&\left[  \boldsymbol{K}_0 +   \lambda_j  \boldsymbol{K}_{\sigma}  \right]\boldsymbol{\phi}_j = \boldsymbol{0}, \\
  \boldsymbol{\phi}_j|_{x=1}=e^{ik_1}\boldsymbol{\phi}_j|_{x=0},  \quad
&\boldsymbol{\phi}_j|_{y=1}=e^{ik_2}\boldsymbol{\phi}_j|_{y=0},\quad  \boldsymbol{\phi}_j|_{z=1}=e^{ik_3}\boldsymbol{\phi}_j|_{z=0}, \nonumber
\end{align}	
where $\boldsymbol{K}_{\sigma} $ is the stress stiffness matrix, the smallest eigenvalue, $\lambda_1$, is the critical buckling strength for the given $\boldsymbol{k}$-vector, $\boldsymbol{k}=[k_1,k_2,k_3]^T$, $i=\sqrt{-1}$ is the imaginary unit and $\boldsymbol{\phi}_1$ is the associated eigenvector. The material buckling strength, $\sigma_c$, is determined by the smallest eigenvalue for all the possible wave-vectors, located in the first  Brillouin zone, $\lambda_{min}$, i.e., $\sigma_c=\lambda_{min}\sigma_1$.  The critical buckling mode is defined as the eigenvector associated with $\lambda_{min}$. The first  Brillouin zone is  the primitive cell in reciprocal space. For a cubic microstructure of unit size, the first Brillouin zone spans over  $k_j \in [-\pi,\ \pi], \ j=1,2,3$.  The first Brillouin zone is further reduced to the irreducible Brillouin zone by the symmetries shared by the microstructure geometry and the macroscopic stress state.  Figure \ref{Fig:IRZ} illustrates the irreducible  Brillouin zone for uni-axial stress state for microstructures with cubic symmetry. Numerical results show that for the considered microstructures, the critical buckling modes lie on the edges of the irreducible Brillouin zone.  Figure \ref{Fig:Geom} shows the buckling strength of all the considered microstructures under the uni-axial stress for  the $\boldsymbol{k}$-vectors along the irreducible Brillouin zone edge. The smallest value for each microstructure represents its buckling strength.

Plate microstructures with volume fractions of $f \in [10^{-4}, 10^{-2}]$ are modelled using 3D shell elements in COMSOL, while truss microstructures are modelled analytically in the low volume fraction limit. Higher volume fraction microstructures are analysed using 3D solid elements in COMSOL. In the two-term interpolation scheme, the first coefficient term is obtained by the slope of the data points from the aforementioned low volume fraction studies.  The coefficient of the second term is obtained by curve fitting in MATLAB by fitting the properties of microstructures with $f \in [10^{-4},\  0.5]$.

\section{Beam model}\label{App:beam}

This Appendix derives the performance equations for the considered beam model for the case of variable height and variable width instead of square cross section discussed in the main text.

\subsubsection*{Variable height ($w$ fixed)}

The mass of the fixed-width beam subject to displacement constraint can be found from (\ref{eq:beamprops})
\be
   m_\delta = \left(\frac{3}{2}\right)^\frac{1}{3} \left(\frac{V L^2}{w \delta^*}\right)^\frac{1}{3} w L \frac{f \rho_0}{E_f^{\frac{1}{3}}} = \left(\frac{3}{2}\right)^\frac{1}{3} \left(\frac{V L^2}{w \delta^*}\right)^\frac{1}{3} w L \frac{1}7{M_2}, \quad
   M_2 = \frac{E_f^{\frac{1}{3}}}{f \rho_0} = \psi_B^\delta \frac{{E_0}^\frac{1}{3}}{\rho_0}, \quad
   \psi_B^\delta = \frac{\tilde{E}_f^\frac{1}{3}}{f}.
   \label{eq:m_delta_h}
\ee
Similarly, the mass of the beam constrained by microstructural buckling failure can also be found from (\ref{eq:beamprops})
\be
   m_c = (6 V w)^{\frac{1}{2}} L \frac{f \rho_0}{\sigma_{c,f}^{\frac{1}{2}}} = (6 V w)^{\frac{1}{2}} L \frac{1}{M_1}, \quad
   M_1 = \frac{\sigma_{c,f}^{\frac{1}{2}}}{f \rho_0} = \psi_B^c \frac{E_0^{\frac{1}{2}}}{\rho_0}, \quad
   \psi_B^c = \frac{{\tilde{\sigma}_{c,f}}^{\frac{1}{2}}}{f},
   \label{eq:m_sigma_h}
\ee
and microstructural yield
\be
   m_y = (6 V w)^{\frac{1}{2}} L \frac{f \rho_0}{\sigma_{y,f}^{\frac{1}{2}}} = (6 V w)^{\frac{1}{2}} L \frac{1}{M_1}, \quad
   M_1 = \frac{\sigma_{y,f}^{\frac{1}{2}}}{f \rho_0} = \psi_B^y \frac{\sigma_0^{\frac{1}{2}}}{\rho_0}, \quad
   \psi_B^y = \frac{\tilde{\sigma}_{y,f}^{\frac{1}{2}}}{f} .
   \label{eq:m_sigma_h}
\ee
In this case, the simple exponents determining advantage of porous microstructure are 3 and 2 for the displacement and failure cases, respectively. This means that worse performing microstructures (higher exponents) than in the square cross section case are advantageous compared to the solid beam. Naturally, mass of the optimal beam will hence also be lower than for the square cross section case.

The coupling factor is found as
\be
   M_2 = 96^{-\frac{1}{6}} \frac{w^\frac{1}{6} L^\frac{2}{3}} {V^\frac{1}{6} (\delta^*)^\frac{1}{3}} M_1 =
         96^{-\frac{1}{6}} \left( \frac{w L^4}{V (\delta^*)^2} \right)^\frac{1}{6} M_1 = C M_1, \quad
         C=96^{-\frac{1}{6}} \left( \frac{w L^4}{V (\delta^*)^2} \right)^\frac{1}{6}.
\ee

\subsubsection*{Variable width ($h$ fixed)}

The mass of the fixed-height beam subject to displacement constraint can be found from (\ref{eq:beamprops})
\be
   m_\delta = \frac{3}{2} \left(\frac{V L^3}{h^2 \delta^*}\right) \frac{f \rho_0}{E_f} = \frac{3}{2} \left(\frac{V L^3}{h^2 \delta^*}\right) \frac{1}{M_2}, \quad
   M_2=\frac{E_f}{f \rho_0} = \psi_B^\delta \frac{E_0}{\rho_0}, \quad
   \psi_B^\delta = \frac{\tilde{E}_f}{f}.
  \label{eq:m_delta_w}
\ee
Similarly, the mass of the beam constrained by microstructural buckling failure can also be found from (\ref{eq:beamprops})
\be
   m_c = \frac{6 V L}{h} \frac{f \rho_0}{\sigma_{c,f}} = \frac{6 V L}{h} \frac{1}{M_1}, \quad
   M_1 = \frac{\sigma_{c,f}}{f \rho_0} = \psi_B^c \frac{E_0}{\rho_0}, \quad
   \psi_B^c = \frac{{\tilde{\sigma}_{c,f}}}{f},
   \label{eq:m_sigma_w}
\ee
and microstructural yield
\be
   m_y = \frac{6 V L}{h} \frac{f \rho_0}{\sigma_{y,f}} = \frac{6 V L}{h} \frac{1}{M_1}, \quad
   M_1 = \frac{\sigma_{y,f}}{f \rho_0} = \psi_B^y \frac{\sigma_0}{\rho_0}, \quad
   \psi_B^y = \frac{\tilde{\sigma}_{y,f}}{f}
   \label{eq:m_sigma_w}
\ee
In this case, the simple exponents determining advantage of porous microstructure are 1 for both displacement and failure cases. This means that it is never advantageous to introduce porosity in the variable width case. Actually, it is a disadvantage because $\psi$ values will always be below one for isotropic or cubic symmetric material microstructures.

The coupling factor is found from
\be
   M_2 = \frac{1}{4} \frac{L^2}{h \delta^*} M_1 = C M_1, \quad C=\frac{1}{4} \frac{L^2}{h \delta^*}.
\ee

\subsection{Column case}
For a column subject to (longitudinal) displacement, buckling and yield constraint, the displacement, stress and buckling load equations are as follows.
\be
   \delta = \frac{P L}{E_f A}, \quad \sigma_{max} = \frac{P}{A}, \quad    P_{k} = \pi^2 \frac{EI}{L^2}.
\ee
For a square cross section $w=h$, these expressions become
\be
   \delta = \frac{P L}{E_f w^2}, \quad \sigma_{max} = \frac{P}{w^2},
   \quad P_k = \frac{\pi^2}{12} \frac{E w^4}{L^2}. \label{eq:stress3}
\ee

Now we can proceed to find masses for the four cases.
Displacement constraint
\be
   m_\delta = \frac{P L^2 f \rho_0}{ \delta^* E_f} = \frac{P L^2}{ \delta^*}\frac{1}{M_2}, \quad M_2 = \frac{E_f}{f \rho_0} = \psi^\delta \frac{E_0}{\rho_0}, \quad \psi^\delta = \frac{\tilde{E}_f}{f}.
\ee
Global buckling constraint
\be
   m_{gc} = \frac{2 \sqrt{3}}{\pi} \frac{L^2 \rho_0  f \sqrt{P}}{\sqrt{E_f}} =
     \frac{2 \sqrt{3} \sqrt{P} L^2}{\pi}\frac{1}{M_1}, \quad
     M_1 =\frac{\sqrt{E_f}}{f \rho_0} = \psi^c \frac{\sqrt{E_0}}{\rho_0}, \quad
     \psi^c = \frac{\sqrt{\tilde{E}_f}}{f}.
\ee
Yield constraint
\be
   m_y = \frac{P f \rho_0 L}{\sigma_{y,f}} = PL \frac{1}{M_1}, \quad M_1 = \frac{\sigma_{y,f}}{f \rho0} = \psi^y \frac{\sigma_0}{\rho0}, \quad \psi^y = \frac{\tilde{\sigma}_y}{f}.
\ee
Local buckling constraint
\be
   m_c = \frac{P f \rho_0 L}{\sigma_{c,f}} = PL \frac{1}{M_1}, \quad M_1 = \frac{\sigma_{c,f}}{f \rho0} = \psi^c \frac{E_0}{\rho0}, \quad \psi^c = \frac{\tilde{\sigma}_c}{f}.
\ee

\end{document}